\documentclass{aa}   
\usepackage[varg]{txfonts}
\usepackage{graphicx}
\usepackage{siunitx}
\usepackage{xcolor}
\usepackage{natbib}
\usepackage{booktabs}
\usepackage{nccmath}
\usepackage{amsmath}
\usepackage{tabularx}
\usepackage{algorithm2e}
\usepackage{csquotes}
\usepackage{comment}
\usepackage{gensymb}
\usepackage{enumitem}
\usepackage{float}
\usepackage{subcaption}
\usepackage[labelfont=bf]{caption}
\usepackage{mwe}
\usepackage[colorlinks,citecolor=blue,urlcolor=blue,bookmarks=false,allcolors=blue,hypertexnames=true]{hyperref} 
\usepackage{threeparttable}
\usepackage{color}
\usepackage[colorinlistoftodos]{todonotes}
\usepackage{tabularx}

\newcommand\clearrow{\global\let\rowmac\relax}

\def \kms         {km$\,$s$^\mathrm{-1}$}

\def \arcmin      {\text{$^\prime$}}
\def \arcsec      {\text{$^\mathrm{\prime\prime}$}}

\def \mjybeam     {mJy\,beam$^\mathrm{-1}$}
\def \mujybeam    {$\mathrm{\mu}$Jy\,beam$^\mathrm{-1}$}

\def \whz         {\,W\,Hz$^{-1}$}
\def \simi        {$\sim$}

\defcitealias{Yun1999}{Y99}
\defcitealias{Liu2019}{L19}
\defcitealias{RodriguezZaurin2014a}{RZ14}
\defcitealias{Spence2016}{S16}

\newlist{inlineroman}{enumerate*}{1}
\setlist[inlineroman]{itemjoin*={{ }},afterlabel=~,label=\roman*)}

\makeatletter
\renewcommand{\fnum@figure}{Figure \thefigure}
\makeatother

\begin{document} 
\title{Seeing the forest and the trees: a radio investigation of the ULIRG Mrk\,273\thanks{Fits files of the radio maps are available at the CDS via anonymous ftp to cdsarc.u-strasbg.fr (130.79.128.5)
or via http://cdsweb.u-strasbg.fr/cgi-bin/qcat?J/A+A/}}
\author{P. Kukreti\inst{1,2}\thanks{\email{kukreti@astro.rug.nl}},
R. Morganti\inst{2,1}, M. Bondi\inst{3}, T. Oosterloo\inst{2,1}, C. Tadhunter\inst{4}, L.K. Morabito\inst{5,6}, E.A.K. Adams\inst{2,1}, B. Adebahr\inst{7}, W.J.G. de Blok\inst{2,8,1}, F. de Gasperin\inst{9,3}, A. Drabent\inst{10}, K.M. Hess\inst{11,2,1}, M.V. Ivashina\inst{12}, A. Kutkin\inst{2,17}, \'A.M. Mika\inst{2}, L.C. Oostrum\inst{2,13,14}, T.W. Shimwell\inst{2,15}, J.M. van der Hulst\inst{1}, J. van Leeuwen\inst{2,13}, R.J. van Weeren\inst{15}, D. Vohl\inst{13,2}, J. Ziemke\inst{2,16}}

\authorrunning{Kukreti et al.}
\titlerunning{A radio investigation of  Mrk\,273}

\institute{
Kapteyn Astronomical Institute, University of Groningen, Postbus 800, 9700 AV Groningen, The Netherlands
\and
ASTRON, the Netherlands Institute for Radio Astronomy, Oude Hoogeveensedijk 4, 7991 PD Dwingeloo, The Netherlands
\and
INAF - Istituto di Radioastronomia, via P. Gobetti 101, 40129, Bologna, Italy
\and
Department of Physics and Astronomy, University of Sheffield, Sheffield S3 7RH, UK
\and
Centre for Extragalactic Astronomy, Department of Physics, Durham University, Durham DH1 3LE, UK
\and
Institute for Computational Cosmology, Department of Physics, University of Durham, South Road, Durham DH1 3LE, UK
\and
Astronomisches Institut der Ruhr-Universit{\"a}t Bochum (AIRUB), Universit{\"a}tsstrasse 150, 44780 Bochum, Germany
\and 
Department of Astronomy, University of Cape Town, Private Bag X3, Rondebosch 7701, South Africa
\and
Hamburger Sternwarte, Universitat Hamburg, Gojenbergsweg 112, 21029, Hamburg, Germany
\and
Th\"uringer Landessternwarte, Sternwarte 5, D-07778 Tautenburg, Germany
\and
Instituto de Astrofísica de Andalucía (CSIC), Glorieta de la Astronomía s/n, 18008 Granada, Spain
\and
Department of Electrical Engineering, Chalmers University of Technology, Gothenburg, Sweden
\and
Anton Pannekoek Institute, University of Amsterdam, Postbus 94249, 1090 GE Amsterdam, The Netherlands
\and
Netherlands eScience Center, Science Park 140, 1098 XG, Amsterdam, The Netherlands
\and
Leiden Observatory, Leiden University, P.O.Box 9513, NL-2300 RA, Leiden, The Netherlands
\and
University of Oslo Center for Information Technology, P.O. Box 1059, 0316 Oslo, Norway
\and
Astro Space Center of Lebedev Physical Institute, Profsoyuznaya Str.\ 84/32, 117997 Moscow, Russia
}

  \abstract{Galaxy mergers have been observed to trigger nuclear activity by feeding gas to the central supermassive black hole. One such class of objects are Ultra Luminous InfraRed Galaxies (ULIRGs), which are mostly late stage major mergers of gas-rich galaxies. Recently, large-scale ($\sim$100\,kpc) radio continuum emission has been detected in a select number of ULIRGs, all of which also harbour powerful Active Galactic Nuclei (AGN). This hints at the presence of large-scale radio emission being evidence for nuclear activity. Exploring the origin of this radio emission and its link to nuclear activity requires high sensitivity multi-frequency data. We present such an analysis of the ULIRG Mrk\,273. Using the International LOFAR telescope (ILT), we detected spectacular large-scale arcs in this system. This detection includes, for the first time, a giant $\sim$190\,kpc arc in the north. We propose these arcs are fuelled by a low power radio AGN triggered by the merger. We also identified a bright $\sim$\,45\,kpc radio ridge, which is likely related to the ionised gas nebula in that region. We combined this with high sensitivity data from APERture Tile In Focus (Apertif) and archival data from the Very Large Array (VLA) to explore the spectral properties. The ILT simultaneously allowed us to probe the nucleus at a resolution of $\sim$0.3\arcsec, where we detected three components, and, for the first time, diffuse emission around these components. Combining this with archival high frequency VLA images of the nucleus allowed us to detect absorption in one component, and a steep spectrum radio AGN in another. We then extrapolate from this case study to the importance of investigating the presence of radio emission in more ULIRGs and what it can tell us about the link between mergers and the presence of radio activity.}


    \date{Received 24 January 2022 / Accepted 04 May 2022}    

    \maketitle
\section{Introduction}
\label{introduction}
Galaxy mergers are spectacular events that are considered to be the way massive galaxies form. They are also thought to be one of the mechanisms that trigger AGNs, including powerful radio AGNs, by helping the gas reach the nuclear regions and accrete onto the central supermassive black hole (SMBH; e.g. \citealt{Hopkins2008,Yuan2010}). Galaxy mergers go through different phases, which can show intense star formation or nuclear activity, and they can be identified using different diagnostics. A class of objects related to a merger phase are UltraLuminous InfraRed Galaxies (ULIRGs; e.g.  \citealt{Sanders1996,Rigopoulou1999,DeLucia2006}), which are understood to be formed by major mergers of gas-rich spiral galaxies. These are extremely bright IR sources with typical luminosities of L$_{\mathrm{IR}}\geq10^{12}$L$_{\odot}$ (for a review see \citealt{Lonsdale2006}). ULIRGs are known to be powered by both starbursts and AGNs (e.g. \citealt{Veilleux2009,Nagar2003,Nardini2010}), and they can therefore represent different phases in the evolutionary sequence of a major merger. Although the properties of optical and radio AGNs in mergers and their feedback effects have been investigated by many studies (e.g. \citealt{Silk1998,RamosAlmeida2011,Cano-Diaz2012,Cresci2015,Carniani2016,Murphy2013,Barcos-Munoz2017}), the presence and properties of large-scale radio emission in such systems are less known.\par

The AGN phase in mergers is understood to occur later than the starbursts. This is due to the fact that in order to be accreted to the central SMBH, the gas has to reach the central few parsecs and lose the majority of its angular momentum \citep{Shlosman1990TheNuclei, DiMatteo2005EnergyGalaxies,Hopkins2006}, whereas the starburst can precede on galactic scales \citep{Mihos1996,Springel2005}. However, optical imaging studies have found that AGNs are also triggered prior to the coalescence of the galaxy nuclei, that is at the pre-coalescence stage \citep{RamosAlmeida2011,Bessiere2012,Pierce2021}. Pierce et al. (in prep) have found that this is true for the majority (61\%) of nearby type 2 quasars. In general, phases of AGN activity are understood to peak either at the pre-coalescence stage or following the coalescence of the merging nuclei \citep{Mihos1996,Tadhunter2011}.\par

Investigating the presence of radio AGNs is interesting because by using the radio spectrum, we can gain information about the timescales on which nuclear activity is triggered in mergers. This will help us understand the role mergers play in triggering radio galaxies. For instance, \citet{Emonts2006} studied the radio properties of B2\,0648+27 and found a significant time delay between the merger-starburst event and the triggering of radio activity. \par

In their study of radio continuum emission from ULIRGs, \citet{Yun1999} (hereafter \citetalias{Yun1999}) estimated that only $\sim$30\% of the mergers host an AGN, on the basis of mid-IR excess. Out of these AGNs, $\sim$40\% host a radio AGN. Among the merger systems with an AGN identified in mid-IR, \citetalias{Yun1999} observed large (hundreds of kiloparsecs) radio continuum tails in a small number of cases (including Mrk\,273). They concluded that the luminosity, size, high degree of polarisation (20$-$50\%), and the presence of a powerful AGN in these systems suggested that the power source for these tails were the AGNs and not the nuclear starburst. Similar radio features have been detected in more ULIRGs recently. \citet{Hayashi2021} discovered radio emission with a size of $\sim$100\,kpc in IRAS\,01004-2237. The two-sided morphology and lack of X-ray detection suggests that the radio emission was not powered by a cluster merger, but an AGN in that system. \citet{Nandi2021} also observed extended radio emission in IRAS\,15001+1433, which hosts an AGN as well. They also conclude that ULIRGs harbour young ($\sim$0.4$-$20\,Myr) radio sources and are possible progenitors of radio galaxies. So these detections beg the question $-$ are such large-scale tails signposts of radio AGN activity in ULIRGs?\par
\begin{figure}
\includegraphics[width=\columnwidth]{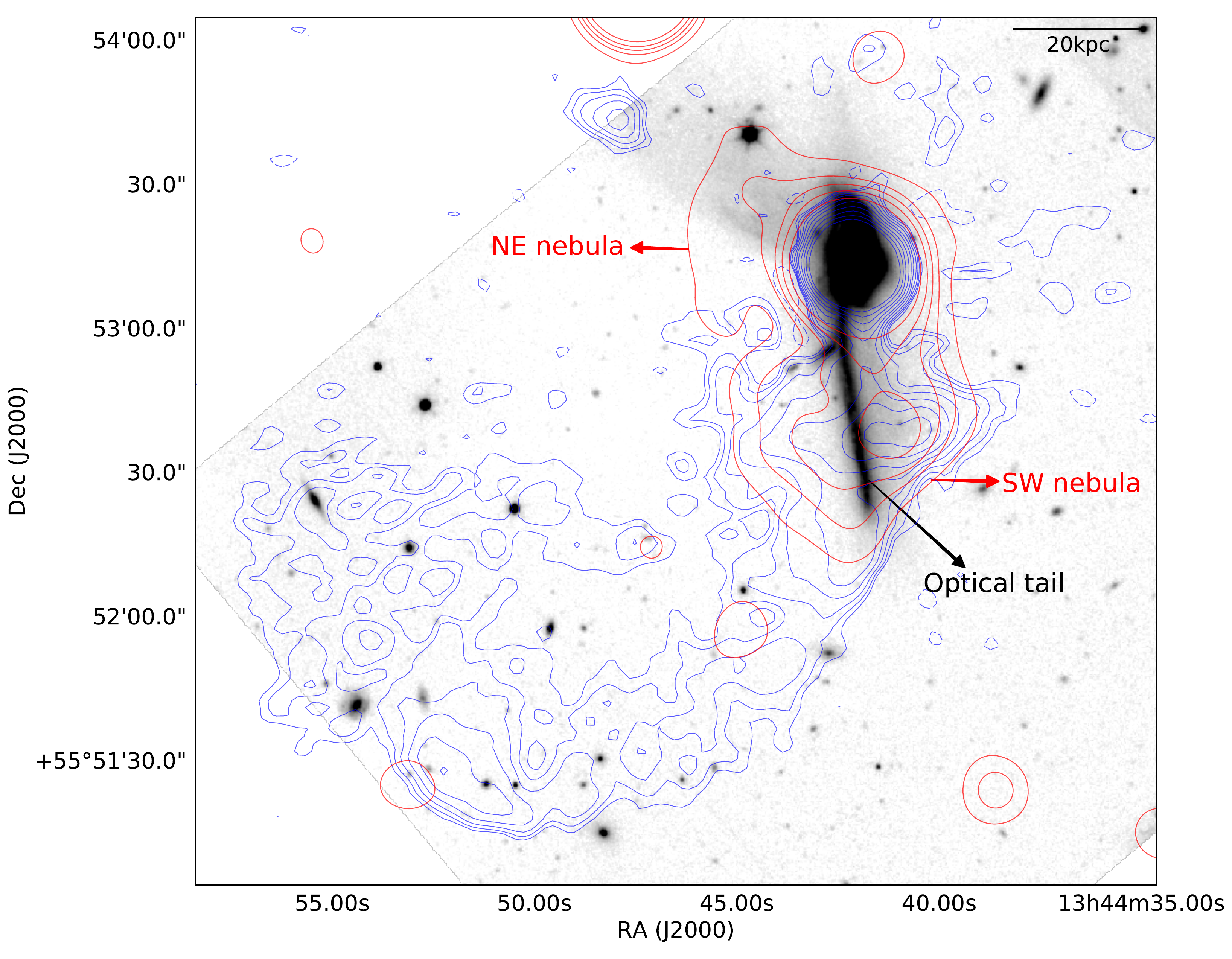}
\caption{Multi-wavelength emission from Mrk\,273. The  blue contours map the 144\,MHz emission from the LOFAR 144\,MHz image, red contours map the soft X-ray emission (0.4$-$2\,keV) from \citetalias{Liu2019}, overlaid on the optical continuum image from \citetalias{Spence2016} at $\lambda$ = 6572\AA. The optical continuum tail seen towards the south of the nucleus is marked. The extended soft X-ray nebulae seen in south-west (SW) and north-east (NE) can be seen as well.  The \textit{Chandra} X-ray image has been smoothed with a Gaussian kernel of $\sigma$=10 pixels, using the \texttt{CIAO} package.}
\label{mlwave}%
\end{figure}
To answer this question, detailed studies of the radio properties of ULIRGs are required. One such study investigated the ULIRG Mrk\,231. Radio emission on scales of parsecs to tens of kiloparsecs was detected in this object \citep{Baum1993a,Ulvestad1999}. \citet{Morganti2016} studied the large southern radio lobe of tens of kiloparsecs size and detected a plateau of emission near the nucleus and poorly collimated bridge structure a few kiloparsecs south of the nucleus (see their Fig.\,6). They concluded that the radio lobe emission could either be due to fuelling of fresh electrons by the central AGN, or in situ acceleration of electrons, or both. \par

Although the current detection rate of large-scale radio emission from ULIRGs is low ($\lesssim$10\%), ongoing and upcoming high sensitivity surveys like the LOFAR LBA Sky Survey (LoLSS; \citealt{DeGasperin2021}), LOFAR Two-metre Sky Survey (LoTSS\footnote{https://doi.org/10.1051/0004-6361/202142484}; \citealt{Shimwell2017,Shimwell2019b}), APERture Tile In Focus (Apertif; \citealt{VanCappellen2021}) and Evolutionary Map of the Universe (EMU; \citealt{Norris2011}) provide a great opportunity to investigate its presence for a large sample. \par
\begin{figure*}
  \includegraphics[width=1.01\textwidth,height=7.8 cm]{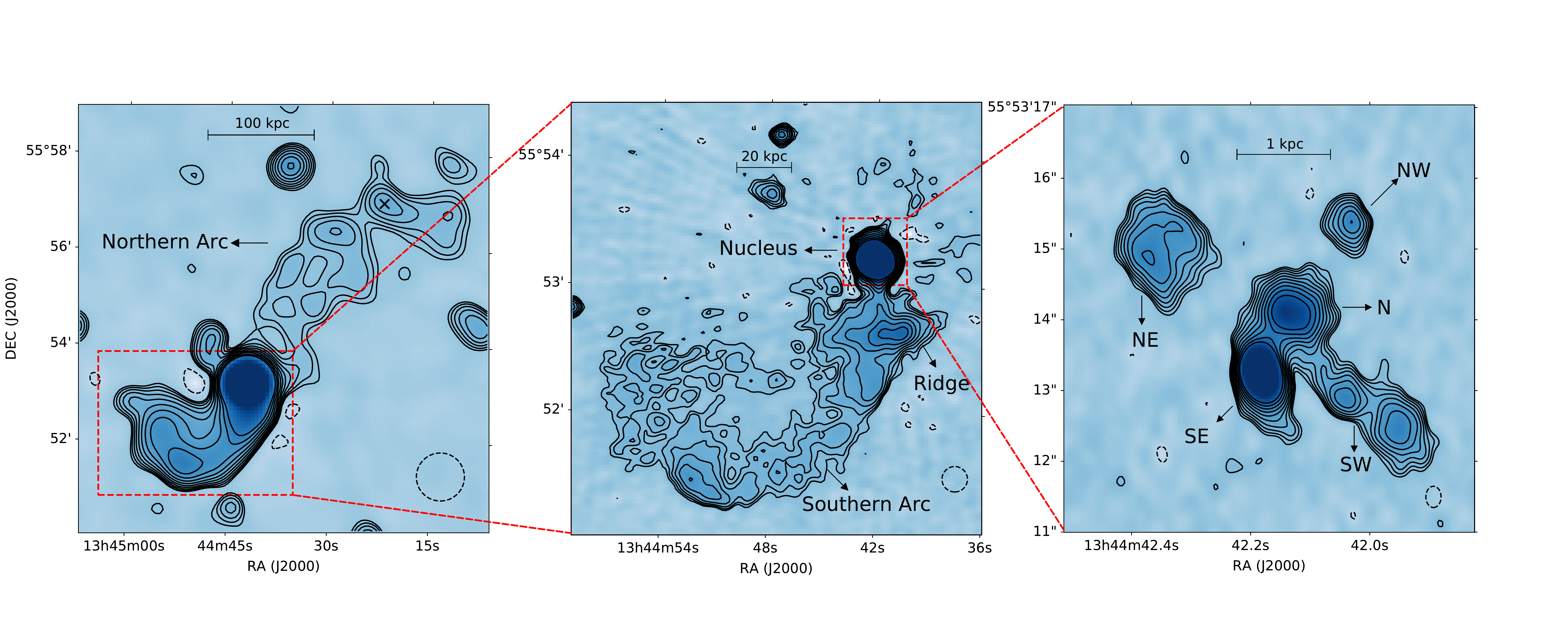}
  \caption{LOFAR 144\,MHz images showing emission at different scales in Mrk\,273. \textit{(left panel)} LoTSS-DR2 image at 30\arcsec~resolution shows the low surface brightness large-scale emission, with the northern arc marked. Black cross in the northern arc marks the nearby galaxy 2MFGC\,11079 at that location. \textit{(middle panel)} LoTSS-DR2 image at 6\arcsec~resolution, showing the southern arc and ridge in Mrk\,273. \textit{(right panel)} Sub-arcsecond image of the nucleus of Mrk\,273 at a resolution of 0.3\arcsec$\times$0.2\arcsec. The northern (N), south-eastern (SE), south-western (SW) and diffuse north-east (NE) components are marked. The contour levels in all images are 3$\sigma_\mathrm{RMS}\times\sqrt{2}^{n}$ where n=0,1,2...10. $\sigma_\mathrm{RMS}$ is 150 \mujybeam, 80 \mujybeam and 90 \mujybeam for the 30\arcsec~image, 6\arcsec~image and sub-arcsecond image, respectively. The negative contours are marked by dashed lines and are at $-3\sigma_\mathrm{RMS}$ level. The beam is drawn with dashed lines in the bottom right corner. Image statistics are summarised in Table~\ref{image_data}.}
  \label{lofar_hba}    
\end{figure*}
In this paper, we present such a study of the merger/ULIRG Mrk\,273 using LOFAR (down to 0.3\arcsec~using international stations) and Apertif. Since this is a well known system, archival data with the Very Large Array (VLA) were also available, which we used to explore the spectral properties. Mrk\,273 is a good object for this study since it is the second nearest ULIRG ($z$~=~0.0373, L$_{\mathrm{IR}}$~=~10$^{12.1}$L$_{\odot}$ ; \citealt{Kim1998a}) and a late stage merger, with a starburst and multiple AGNs. Previous studies have found evidence for three nuclei in the system, two in the near-IR \citep{Knapen1997} $-$ northern (N) and south-western (SW) $-$ and a third south-eastern (SE) in the radio \citep{Condon1991} and [OIII] (\citealt{RodriguezZaurin2014a}, hereafter \citetalias{RodriguezZaurin2014a}). The presence of an AGN in these nuclei has been confirmed in the N component by \citet{Knapen1997} and \citet{Liu2019} (hereafter \citetalias{Liu2019}), in the SW by \citet{Iwasawa2011} and \citetalias{Liu2019}, and a third AGN has been suggested in the SE component by \citetalias{RodriguezZaurin2014a}. \par
The presence of a dense neutral and molecular gas disk \citep{Condon1991,Cole1999,Carilli2000} and 
a collection of compact luminous supernovae remnants and radio supernovae \citep{Bondi2005} have also been confirmed in the N component. \citet{Vardoulaki2015} had suggested the presence of a radio AGN in the nuclear region, on the basis of the steep spectral index ($\alpha\sim1.1$) from 1400$-$8415\,MHz, but they could not pin down its location to one of the three nuclear components. The nuclear region also has a high star formation rate (SFR) of $\sim$139\,M$_{\odot}$\,yr$^{-1}$ \citep{Cicone2014MassiveObservations}. 

The stellar population in the nuclear region and optical tail has an age of 0.7$-$2.0\,Gyr, but there is a very young stellar population of age $\lesssim$50\,Myr present as well \citep{RodriguezZaurin2009}. 
The gas kinematics in the nuclear region of Mrk\,273 is complex, and emission line studies have revealed ionised gas outflows up to $\sim$5\,kpc (\citealt{Rupke2013}; \citetalias{RodriguezZaurin2014a}; \citealt{Spence2016}--hereafter \citetalias{Spence2016}; \citealt{Leung2021a}). On the kiloparsec scales, optical images of the system show a $\sim$30\,kpc tidal tail, filaments, and diffuse emission to the south and north-east of the nucleus (\citetalias{RodriguezZaurin2014a,Spence2016}, also see Fig.~\ref{mlwave}). \par

Although large-scale radio emission in Mrk\,273 was reported by \citetalias{Yun1999}, their low resolution ($\sim$20\arcsec) did not allow them to resolve the different features of this emission. Using LOFAR, Apertif and VLA data, we have identified extended low surface brightness features in Mrk\,273, and investigated their spectral properties. We have also studied the radio emission from the resolved nuclear components. The paper is structured in the following manner: In Sec 2, we describe the data reduction and the procedure to make the spectral index maps; in Sec 3 we present the morphology and spectral properties of the source, the spectral age estimates and the absorption analysis; and in Sec 4, we discuss the origin of the radio emission and the spectral index trends of the system. Throughout the paper, we define the spectral index $\alpha$ as: S $\propto\nu^{-\alpha}$. The cosmology adopted in this work assumes a flat universe with H$_\mathrm{0}$ = 71 km s$^\mathrm{-1}$ Mpc$^\mathrm{-1}$, $\Omega_\mathrm{m}=0.27$ and $\Omega_\mathrm{vac}=0.73$. At the redshift of Mrk\,273, 1\arcsec~corresponds to 0.73\,kpc.

\section{Data reduction}

\begin{table*}
\centering
\begin{threeparttable}
\caption[]{Image statistics summary.}
         \label{image_data}
\begin{tabular}{cccccc}
            \hline
            \hline
    \noalign{\smallskip}
    Frequency & Telescope  & Resolution (BPA) & RMS noise  & Integrated flux density $\pm$ error\\
    (MHz) & & & (\mujybeam) & (mJy)\\
    \noalign{\smallskip}
    \hline
    \hline
    \noalign{\smallskip}
   
    54 & LOFAR LBA Dutch & 15\arcsec$\times$15\arcsec~ (90\degree) & 1600 & 622$\pm$94 \\
    144 & LOFAR HBA Dutch & 6\arcsec$\times$6\arcsec~ (90\degree) & 80 & 466$\pm$47 \\   
    1360 & Apertif & 16\arcsec$\times$12\arcsec~ (0\degree) & 35 & 154$\pm$8 \\        
    1455 & VLA C configuration & 16.8\arcsec$\times$13\arcsec~(15\degree) & 60 & 159$\pm$8 \\
         \vspace*{10px}
    4700 & VLA D configuration & 18.6\arcsec$\times$13.2\arcsec~(63\degree) & 28 & 80$\pm$4 \\
    144 & LOFAR HBA international & 0.2\arcsec$\times$0.3\arcsec~(-2\degree) & 90 & 258$\pm$26 \\
    4700 & VLA A configuration & 0.45\arcsec$\times$0.38\arcsec~(24\degree) & 30 & 62$\pm$3 \\
    8415 & VLA C configuration & 0.30\arcsec$\times$0.30\arcsec~(0\degree) & 28 & 43$\pm$2 \\
            \hline
            \hline
 \end{tabular}      
    \small{\textbf{Note.} Image statistics of the low resolution images are summarised in the first five rows. The last three rows summarise measurements from the high resolution image of the nucleus of Mrk\,273.}
\end{threeparttable} 
\end{table*}
\label{data reduction}

We have combined data from the International LOFAR Telescope (ILT) Low Band Antenna (LBA) at 42$-$68\,MHz and High Band Antenna (HBA) at 120$-$168\,MHz, Apertif at 1.36\,GHz and VLA at 1.4\,GHz and 4.7\,GHz to trace the radio continuum and spectral properties of Mrk\,273 at both high and low resolutions.

\subsection{LOFAR 144 MHz}
To trace the low frequency radio continuum emission, we have used the mosaic P205+55 from the LOFAR Two-metre Sky Survey (LoTSS; \citealt{Shimwell2019b}).  Mrk\,273 lies at a distance of 1.2\degree~from the phase centre. The dataset is part of LoTSS-DR2 (\citealt{Shimwell2022}\footnote{https://doi.org/10.1051/0004-6361/202142484}) and the observations  were carried out with the standard LoTSS survey setup, i.e. 8 hr on-source observations bookended on either side with a flux density calibrator, in this case 3C\,147. A total of 62 Dutch stations were used for this observation, which included 48 core stations with baselines up to 4 km and 14 remote stations with baselines up to 120 km. This dataset did not have all 13 international stations, which have baselines up to $\sim$2000 km. Therefore, for the sub-arcsecond image, we used data from \cite{Morabito2021}. They used data from a re-observation of this field. This re-observation also used the standard LoTSS observational setup, and the same flux density calibrator but has 13 international stations.\par
Both datasets were recorded with an integration time of 1s, a 48\,MHz bandwidth centred at 144\,MHz and a channel width of 3.05\,kHz. The data were then passed through the standard LOFAR pre-processing pipeline \citep{Heald2010a} which performed the RFI flagging using the AOflagger \citep{Offringa2010Post-correlationMethods,Offringa2012ADetection} and averaged down to a channel width of 12.2\,kHz. Direction independent calibration was then performed using the PREFACTOR\footnote{\url{https://github.com/lofar-astron/prefactor}} pipeline \citep{DeGasperin2019b,VanWeeren2016,Williams2016LOFARCounts}.

\subsubsection{Dutch array image}
The direction dependent self-calibration of the low-resolution Dutch array image was done using the DDF-pipeline, described in \cite{Shimwell2019b,Tasse2020a}. The LoTSS-DR2 data reduction is described in detail in \citet{Shimwell2022}. We have used images of the field at both 6\arcsec~and 20\arcsec~resolution to trace as much of the faint large-scale emission as possible while also being able to map the details in the large-scale structure. The 6\arcsec~image has an RMS noise of 80\mujybeam and an integrated flux density of 466$\pm$47\,mJy measured using 3$\sigma$ contours as reference and is shown in Fig.~\ref{lofar_hba}. The integrated flux density in the 20\arcsec~image is 451$\pm$45\,mJy. The error associated to the measured flux densities is dominated by the flux calibration uncertainty, that is typically assumed to be 10\%. The residual image was noise like - i.e. the object was full deconvolved. To recover more of the extended structure in this object, we have also smoothed the 20\arcsec~image to a resolution of 30\arcsec, as shown in Fig.~\ref{lofar_hba}.  
We recover the same integrated flux density in the 6\arcsec, 20\arcsec~and 30\arcsec~image, within the errors. However, the structure of the northern arc is most clearly visible in the 30\arcsec~image.
\subsubsection{Sub-arcsecond International array image}
The calibration of the international stations data for this field was done using the LOFAR long baseline pipeline\footnote{\url{https://github.com/lmorabit/lofar-vlbi}}. The pipeline and the data reduction for the LoTSS pointing P205+55 are described in detail using the same field in \cite{Morabito2021}. The paper also described the procedure for the delay and self-calibration of the delay calibrators in the field \citep{Jackson2016,Jackson2021}. 
After applying the self-calibration solutions from the delay calibrator to the target, we have used the procedure from \citet{VanWeeren2021} to iteratively self-calibrate the phases and amplitudes and finally make the final image of the target. This procedure uses WSClean \citep{Offringa2014Wsclean:Astronomy,Offringa2017AnImages}. Using the Briggs weighting scheme and a \texttt{robust = $-$1} along with a minimum uv length of 40k$\lambda$, we obtain a resolution of 0.2\arcsec$\times$0.3\arcsec~and an RMS noise of 90 \mujybeam for the target image, see Table~\ref{image_data}. The image is shown in Fig.~\ref{lofar_hba}. To recover more extended emission around the nuclear components, we made another image with \texttt{robust = 0} and a minimum uv length of 2k$\lambda$. This image has a resolution of 0.4\arcsec$\times$0.5\arcsec, an RMS noise of 78 \mujybeam and is shown in Fig.~\ref{lofar_highres}.

\subsection{LOFAR 54 MHz}
We have used the image from the mosaic P205+55 of the LOFAR LBA Sky Survey (LoLSS; \citealt{DeGasperin2021}). The survey observations were carried out with the LOFAR Low Band Antenna (LBA) Dutch stations for $\sim$8\,hr and a 24\,MHz bandwidth centred at 54\,MHz. The time and frequency resolution were 1s and 3.052\,kHz before Radio Frequency Interference (RFI) flagging, and were averaged to 2s and 48.828\,kHz after flagging. The data reduction is described in detail in the survey paper and in \citet{DeGasperin2020}. The image resolution was 15\arcsec~with an RMS noise of $\sim$1.6 \mjybeam and is shown in Fig.~\ref{lba}.

\subsection{Apertif 1360\,MHz}
The APERture Tile In Focus (Apertif) phased-array feed (PAF) system is installed at the Westerbork Synthesis Radio Telescope (WSRT) and operates around 1400~MHz (see \citealt{VanCappellen2021} for the full description). Because of the large field of view (the total area covered in one observation by such a mosaic is about $3.5^\circ \times 3^\circ$) combined with the sensitivity of the radio continuum images (about 30$-$40 $\mu$Jy) and the spatial resolution of about $12^{\prime\prime} \times 12^{\prime\prime}/\sin\delta$, Apertif is an ideal survey instrument and provides an ideal complement to the surveys done with LOFAR.\par
The Apertif imaging surveys started on 1 July, 2019, with the aim of covering up to $\sim$2500 square degrees of the northern sky. The first data from the Apertif surveys---providing radio continuum, HI, and polarisation images and cubes---have become publicly available in the first data-release, which was done in November 2020 \footnote{http://hdl.handle.net/21.12136/B014022C-978B-40F6-96C6-1A3B1F4A3DB0}. The observations covering the position of Mrk\,273 were taken on the 22 December 2020 (ObsID 201222016) with a total observing time of 11 h. 3C\,196 was used as the flux calibrator. The data of the beams including Mrk\,273 were flagged and cross calibrated by the automatic Apertif pipeline, Apercal \citep{Adebahr2021}. The self-calibration and the final image were made manually using MIRIAD \citep{Sault1995}, followed by the primary beam correction (Kutkin et al. in prep\footnote{https://github.com/akutkin/amosaic}). The image is shown in Fig.~\ref{apertif} and has an RMS noise of 35 $\mu$Jy beam$^{-1}$ and a restoring beam of 12\arcsec$\times$16\arcsec~(PA = 0$\degree$), see Table~\ref{image_data}.

\subsection{Archival data}

\subsubsection{Low resolution VLA 1.45\,GHz and 4.7\,GHz}
\label{low res high freq}
We reprocessed archival VLA 1.45\,GHz and 4.7\,GHz observations of the target. 1.45 GHz observations were carried on 31 October 2010 with the VLA in C configuration (Project code: 10B-212). The target was observed for $\sim$50 mins with an integration time of 1s and a 128\,MHz bandwidth split into 64 channels of 2\,MHz each. 3C\,286 and J1349+5341 were used as the flux and phase calibrators respectively.\par
Mrk\,273 was also observed at 4.7 GHz on 29 August 2010 with the VLA in D configuration (Project code: AL746). The target was observed for $\sim$6 mins with an integration time of 1s and a bandwidth of 1024\,MHz. The entire bandwidth was split into 8 equal width sub-bands of 128\,MHz, with a sub-band split into 64 channels of 2\,MHz each. 3C\,286 was used as the flux calibrator and J1339+6328 \& J1349+5341 were used as the phase calibrators.\par  
The data were manually flagged and cross-calibrated in Astronomical Image Processing System (AIPS, \citealt{Greisen2003}). The flux scale was set according to \citet{Perley2013}. Flagging on the C band data was done using the task \texttt{FLAGR} in AIPS. This data was then self-calibrated and imaged using Common Astronomy Software Applications (CASA, \citealt{McMullin2007}). The final images were made using Briggs weighting with a \texttt{robust = 0.5} for the 1.45 \,GHz and 4.7\,GHz image. Self-calibration of the dataset was done initially with phase only self-calibration, followed by amplitude and phase self-calibration. The resolution obtained was 16.8\arcsec$\times$13\arcsec~(PA = 15\degree) for the 1.45\,GHz image and 18.6\arcsec$\times$13.2\arcsec~(PA = 63\degree) for the 4.7\,GHz image. The RMS noise is 60 \mujybeam for the 1.45\,GHz image and 28 \mujybeam for the 4.7\,GHz image, see Fig.~\ref{continuum_lowres} and Table~\ref{image_data}. Primary beam correction was done for both images using the task \texttt{impbcor} in CASA.

\subsubsection{High resolution VLA 4.7\,GHz and 8.4\,GHz}
We have reprocessed VLA 4.7\,GHz data to image the nucleus of Mrk\,273 and match the spatial resolution of the LOFAR international stations data. Mrk\,273 was observed with the VLA on 22 June 2011 in A configuration (Project code: AL746). The target was observed for $\sim$8 mins with an integration time of 1s and a bandwidth of 1024\,MHz split into 8 equal width sub-bands. 3C\,286 and J1349+5341 were used as the flux density calibrator and phase calibrator respectively. Similar to the low resolution VLA images, the data was flagged, cross-calibrated and bandpass calibrated in AIPS. This data was then self-calibrated and imaged in CASA with the Briggs weighting scheme and a \texttt{robust = 0.5}. For the self-calibration, phase only calibration was performed first, followed by amplitude and phase calibration. Primary beam correction was again done for the image using the task \texttt{impbcor}. The final resolution obtained was 0.45\arcsec$\times$0.38\arcsec~and an RMS noise of 30\mujybeam. \par
We have also used the 8.4\,GHz image from \cite{Condon1991}, made with the VLA in A configuration. The image has a resolution of 0.3\arcsec$\times$0.3\arcsec~and an RMS noise of 28\mujybeam. The images are shown in Fig.~\ref{continuum_lowres} and image statistics are summarised in Table~\ref{image_data}.

\begin{figure}
\includegraphics[width=\columnwidth]{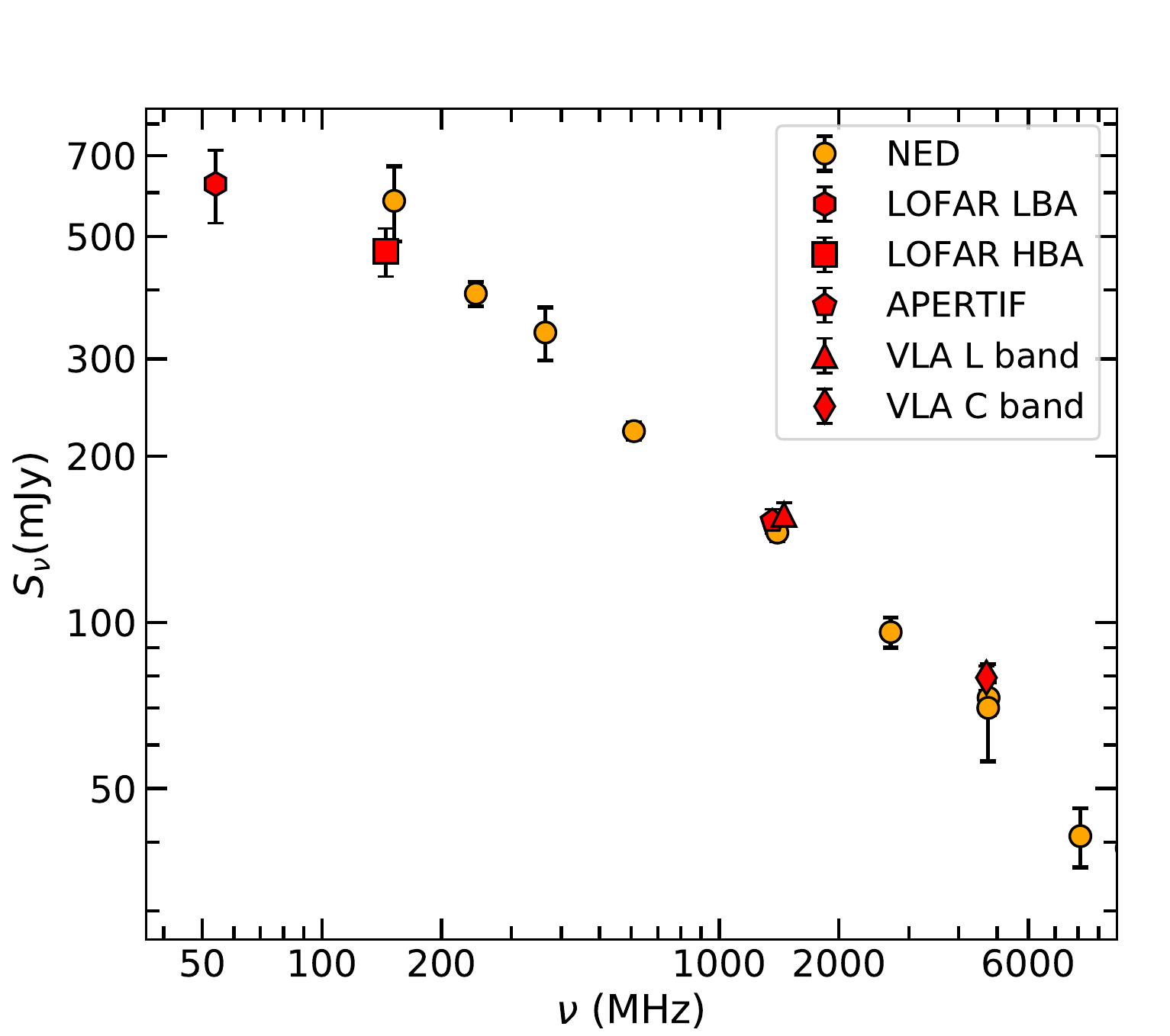}
\caption{Integrated flux densities of Mrk\,273 from our low resolution images, measured using 3\,$\sigma$ contours for reference and NED. The red points show measurements from this study and orange points mark the NED values.}
\label{int_spec}%
\end{figure}

\subsection{Flux density scale}
An accurate flux density scale is necessary for studying the spectral properties of the emission. To check the accuracy of the flux scale of our low resolution images, we have plotted the integrated flux densities of Mrk\,273 along with the values from NED. In our low resolution images, the integrated flux densities were measured using 3$\sigma_\mathrm{RMS}$ contours as reference. We used a flux density scale error of 15\% for LOFAR LBA, 10\% for LOFAR HBA, and 5\% for Apertif 1.36\,GHz, VLA 1.45\,GHz, 4.7\,GHz and 8.4\,GHz data. The plot is shown in Fig.~\ref{int_spec} and shows that our flux scale is in agreement with the literature. Our 144\,MHz flux density is lower, although in agreement within errors with the literature value at 150\,MHz, which is taken from the 6C survey \citep{Hales1990}. This is likely due to the fact that 6C survey has a much lower resolution of $\sim$4.5\arcmin, and therefore collects all the low level flux also from the companions that we excluded.
 \begin{figure*}
        \centering
        \begin{subfigure}[b]{0.5\textwidth}
            \centering
            \includegraphics[width=1\textwidth,height=0.86\textwidth]{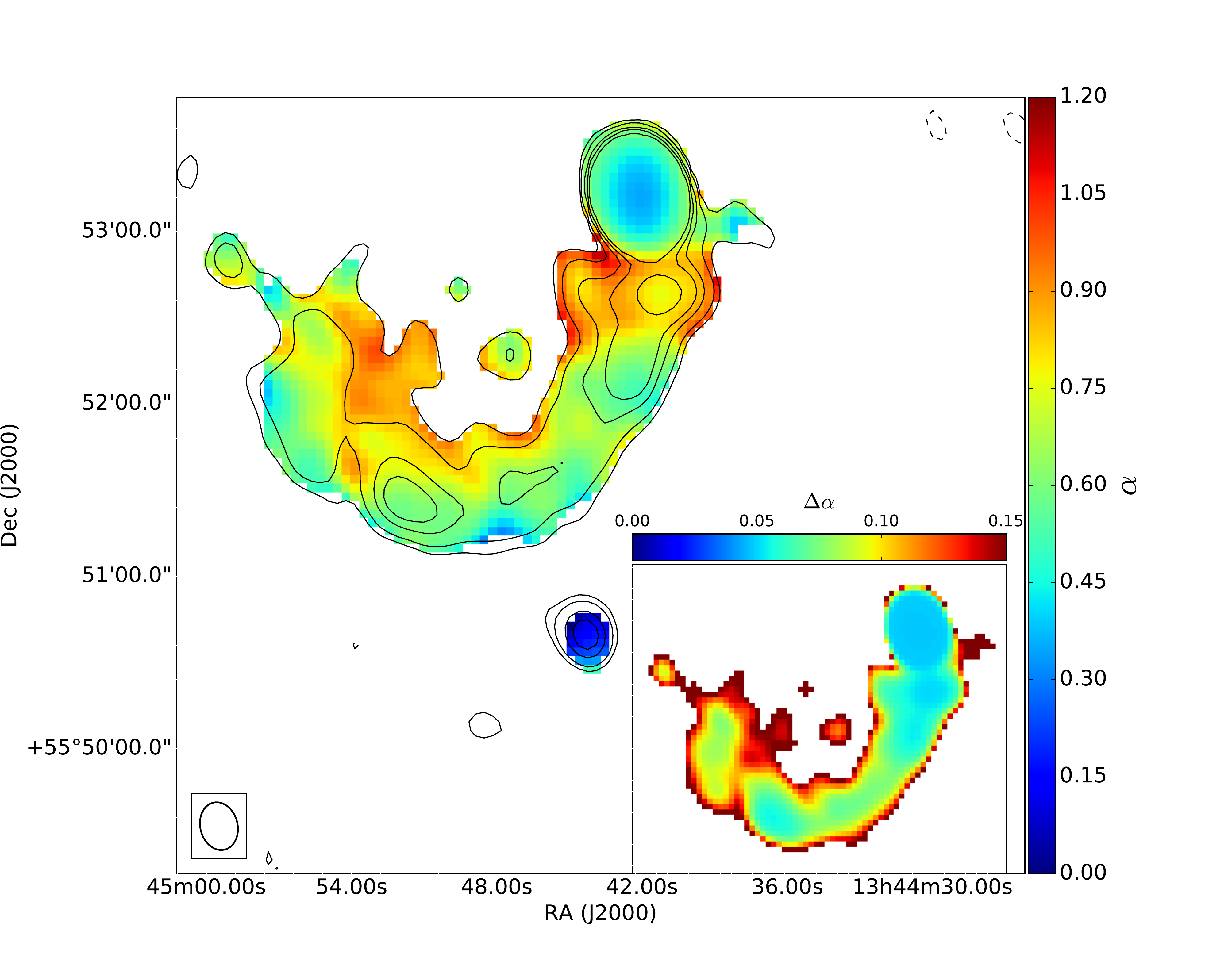}
            \caption[1]%
            {Spectral index map $\alpha^{144}_{1455}$ at 16.8\arcsec$\times$13\arcsec~resolution}    
            \label{spix_six}
        \end{subfigure}
        \begin{subfigure}[b]{0.49\textwidth}  
            \centering 
            \includegraphics[width=1.03\textwidth,height=0.86\textwidth]{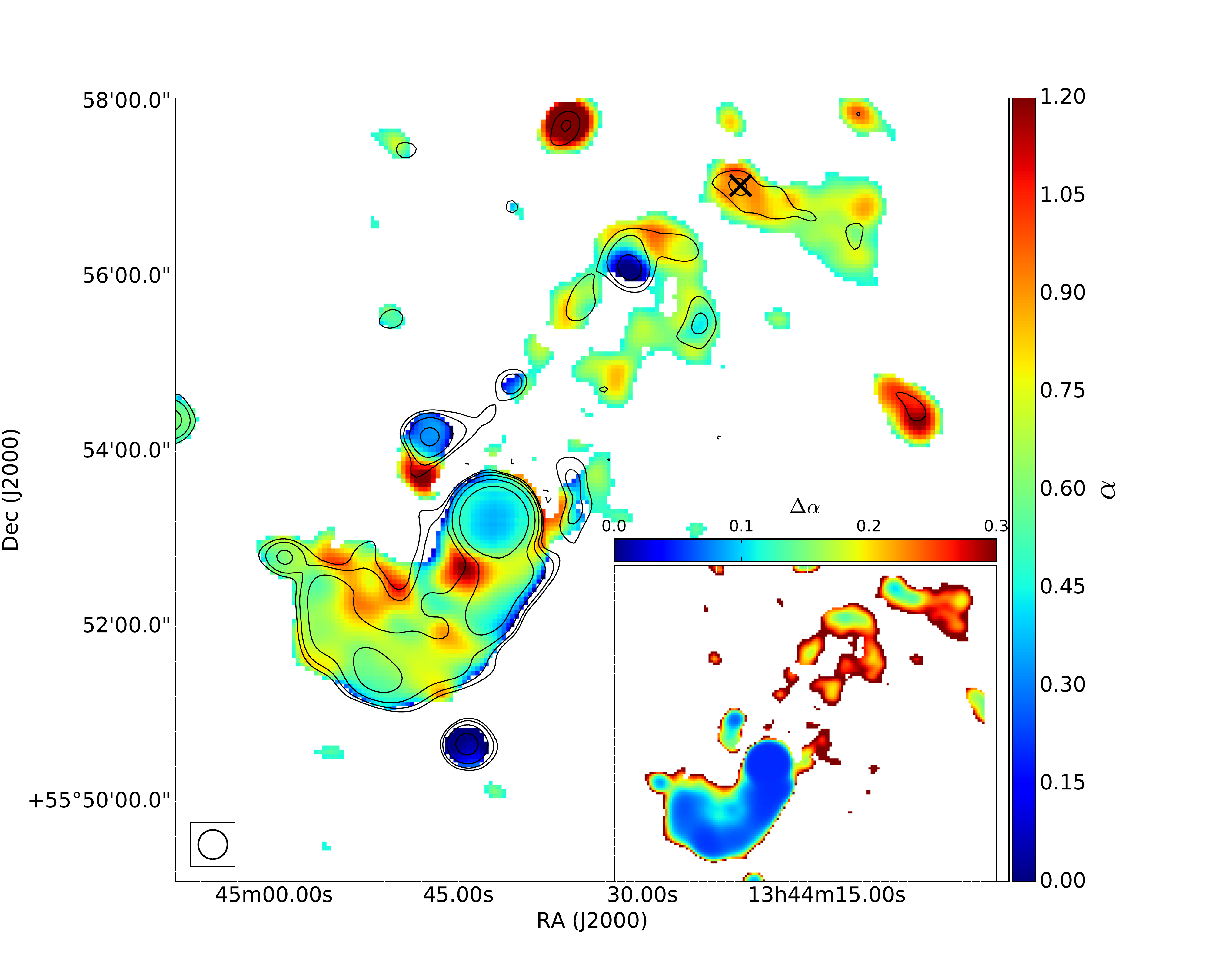}
            \caption[2]%
            {Spectral index map $\alpha^{144}_{1360}$ at 20\arcsec$\times$20\arcsec~resolution}  
            \label{spix_twenty}
        \end{subfigure}     
        
        \caption[international]   
        {\small Spectral index maps of large scale emission in Mrk\,273. (a) Spectral index map from 144$-$1455\,MHz made with the LOFAR and VLA images at 16.8\arcsec$\times$13\arcsec~resolution. The local RMS noise, $\sigma_{RMS}$, in the 144\,MHz image is 200 \mujybeam and in the 1455\,MHz image is 60 \mujybeam. The contours mark the 1455\,MHz emission at levels (-3, 3, 5, 10, 15, 25)$\times\sigma_{RMS}$. The map shows a spectral gradient across the arc, with $\alpha$ $\sim$ 0.6 at the southern outer edge and $\sim$0.9 at the northern inner edge. The ridge has a steep spectral index of $\alpha\approx$ 0.88$\pm$0.07. (b) The 20\arcsec~spectral index map from 144$-$1360\,MHz made with the LOFAR and Apertif images. The local RMS noise, $\sigma_{RMS}$, in the 144\,MHz image is 100 \mujybeam and in the 1360\,MHz image is 37 \mujybeam. The contours mark the 1360\,MHz emission at levels (-3, 3, 5, 10, 25, 100)$\times\sigma_{RMS}$. The pixels outside the 3$\sigma$ contours show the lower limit of the spectral index. The black cross marks the background galaxy present at the location of the northern arc. The inset maps show the spectral index errors.}
        \label{spix}
    \end{figure*}

\section{Results}
\label{results}
We describe here the morphology of the large ($\sim$300\,kpc) and small ($\sim$4\,kpc) scale emission in Mrk\,273, followed by the spectral properties and the estimation of the spectral age of the emission.

\subsection{Morphology}
Mrk\,273 shows interesting morphology on both small and large-scales which we describe below. 

\subsubsection{Nuclear region}
\label{nuclear_morphology}
The radio morphology of the Mrk\,273 nucleus has been observed before at GHz frequencies with sub-arcsecond resolution \citep{Cole1999,Bondi2005} and multiple components were identified. As mentioned before, dual AGNs have been identified in the X-ray in the N and SW component, and another AGN has been proposed in the SE component. Our new 144\,MHz sub-arcsecond image is shown in Fig.~\ref{lofar_hba}. Two bright components, N and SE, of $\sim$0.9\,kpc size can be seen, separated by $\sim$0.7\,kpc. Two bright knots of emission $\sim$0.6\,kpc apart can also be seen for the first time in the SW component, along with some diffuse emission. We also observe diffuse emission of about $\sim$1\,kpc and $\sim$0.4\,kpc in the north-east (NE) and north-west (NW). In our long baseline image at 144\,MHz with a lower minimum uv length and a 0.5\arcsec~resolution (Fig.~\ref{lofar_highres}), we also detect diffuse emission around the components. The interesting feature to note in the figure is the emission extending to $\sim$1.5\,kpc southwards of the SE component, possibly linking it to the southern arc or ridge emission. Some more diffuse emission has been detected around the SW component. The N, SE and SW component had been detected before at 8.4\,GHz by \citep{Condon1991}. However, at 144\,MHz, the SE component is the brightest whereas the N component is brightest in the 8.4\,GHz image. This suggests that the emission in the N component is absorbed at low frequencies. We measured the flux density of each component by fitting a 2D Gaussian over a region of the size of3$\sigma_{RMS}$ level contours at every frequency, where $\sigma_{RMS}$ is the local RMS noise in the image. The flux densities, luminosities and spectral indices of these components are listed in Table~\ref{components}. 

\subsubsection{Large-scale structure}
Fig.~\ref{lofar_hba} shows the large-scale emission in Mrk\,273 at 144\,MHz. The 6\arcsec~map shows the southern arc with a remarkably smooth curvature. This arc has a surface brightness ranging from 0.15$-$0.30\,mJy\,arcsec$^{-2}$ and a total projected linear extent of \simi{100\,kpc} (measured using 3$\sigma$ contours as reference). The emission is slightly more diffuse at the end of the arc than near the nucleus. Although the giant radio continuum 'plume' had been reported before by \citetalias{Yun1999}, this is the first time this large-scale structure has been resolved and other features identified. A new feature we observe is the large arc in the north, shown in the 30\arcsec~image in Fig.~\ref{lofar_hba}. This arc has a much lower surface brightness than its counterpart in the south, 0.04$-$0.08\,mJy\,arcsec$^{-2}$, and a larger projected linear extent of $\sim$190\,kpc. We recover part of this emission in the Apertif image as well, however, we do not have enough sensitivity to recover this emission in the VLA 1455\,MHz and 4700\,MHz images.\par
Another new feature of interest is the bright ridge of emission, located at $\sim$25\,kpc southwards from the nucleus, marked in Fig.~\ref{lofar_hba}.  It is oriented almost perpendicular to the southern optical continuum tail, shown in Fig.~\ref{mlwave}, and is also coincident with the linear feature branching out of the optical tail to the east. It overlaps well with the ionised gas \citepalias{RodriguezZaurin2014a,Spence2016} and soft X-ray (\citealt{Xia2001}; \citetalias{Liu2019}) nebula in this region. The ridge has a surface brightness ranging from 0.5$-$0.7\,mJy\,arcsec$^{-2}$ and a projected linear extent of $\sim$45\,kpc. No radio emission is seen to the north-east of the nucleus in the low resolution images, where extended soft X-ray emission can be seen in Fig.~\ref{mlwave}.
\subsection{Spectral index maps}
In order to trace the large-scale spectral properties in Mrk\,273, we have constructed spectral index maps from 144$-$1455\,MHz at $\sim$16\arcsec~and 20\arcsec, and 1455$-$4700\,MHz at 20\arcsec. We do not construct spectral index maps of the nucleus with our sub-arcsecond images. This is due to the fact that the 1400\,MHz image from \citet{Cole1999} is not publicly available in FITS format. This image is crucial to cover the frequency range from 144\,MHz to 4700\,MHz. Without the 1400\,MHz image, making a spectral index map from 144-4700 MHz would not give an accurate picture of the spectral properties. For the low resolution spectral index maps, we have not used the 54\,MHz image since it did not recover the extended emission as well as the higher frequency images. Spectral index maps require the images to recover emission on the same angular scales and have the same resolution. We have ensured that the data sets used for the low resolution images had short baselines to be sensitive to emission at an angular scale of $\sim$150\arcsec~which is the size of the southern arc (an interferometer with a shortest baseline of D$_{min}$ is sensitive to a largest angular scale of 0.6\,$\lambda$/D$_{min}$; \citealt{Tamhane2015}). To have the same resolution in both images for the spectral index map, the image with higher resolution was smoothed using a 2D Gaussian in the image plane with the task \texttt{IMSMOOTH} in CASA. To trace the spectrum of the northern arc, we also constructed a low resolution spectral index map from 144$-$1360\,MHz after smoothing the Apertif 1360\,MHz image to 20\arcsec~resolution. 
\par

Position offsets in images can be caused by phase calibration errors, and it is necessary to correct for such offsets and align the images before making the spectral index maps. For this purpose, we have fitted a 2D Gaussian over the nucleus of Mrk\,273 in the images and derived the pixel position coordinates of the peak flux density. All images were then regridded and aligned using the position from one image as reference with the tasks \texttt{IMHEAD} and \texttt{IMREGRID} in CASA. The position offset after this procedure was $\leq$0.01 pixels, which is sufficient for our analysis. The spectral index maps were made with the task \texttt{IMMATH} in CASA. Throughout the paper, spectral index errors are calculated as
\begin{ceqn}
\begin{align}
      \Delta\alpha =
          \frac{1}{\ln(\frac{\nu_\mathrm{1}}{\nu_\mathrm{2}})}\sqrt{\left(\frac{\Delta S_\mathrm{1}}{S_\mathrm{1}}\right)^\mathrm{2}+\left(\frac{\Delta S_\mathrm{2}}{S_\mathrm{2}}\right)^\mathrm{2}},
\end{align}
\end{ceqn}

\noindent where $\Delta S_\mathrm{1}$ and $\Delta S_\mathrm{2}$ are errors in the flux densities which include statistical errors in the measurements as well as uncertainties in the overall scale.\par
To derive the spectral properties from 144$-$1400 MHz, we follow two approaches for the southern and northern arc regions. 
The analysis of the southern arc was done done using the spectral index image obtained at $\sim$16\arcsec. The main properties (that will be described in Sec~\ref{large scale emission}) were first derived using the Apertif image and then refined using the VLA image. The spectral index values of the southern arc and the trends in spectral properties with the two were found to be in agreement, and in the remainder of the paper the analysis will be done using the image obtained with the VLA (Fig.~\ref{spix_six}) because it is less affected by calibration artefacts. For this map, emission within the 3$\sigma$ contours was used. \par

In the northern arc region, the high sensitivity to low surface brightness emission of the Apertif image allows it to recover more emission than the VLA 1455\,MHz image, and we use it to obtain the spectral indices in (at least part of) this region (see Sec~\ref{large scale emission}). We smoothed the Apertif map to match the LOFAR 20\arcsec~image and used $>$3$\sigma$ emission to construct a spectral index map from 144$-$1360\,MHz (Fig.~\ref{spix_twenty}). Since the LOFAR image recovered more emission in the northern arc than the Apertif image, we estimate a lower limit of the spectral index for pixels outside 3$\sigma$ contours in the Apertif image. To do this, we have used a value of 3$\sigma$ as the upper limit on the flux density for these pixels. \par

\par
For the nucleus, we have used the integrated flux densities of the different components (see Fig.~\ref{lofar_hba}) from the sub-arcsecond images, to calculate the spectral indices. This is done because the high frequency sub-arcsecond images do not recover all the emission we see in the 144\,MHz sub-arcsecond images. \par

\subsection{Spectral properties}
\begin{figure}
\includegraphics[width=\columnwidth]{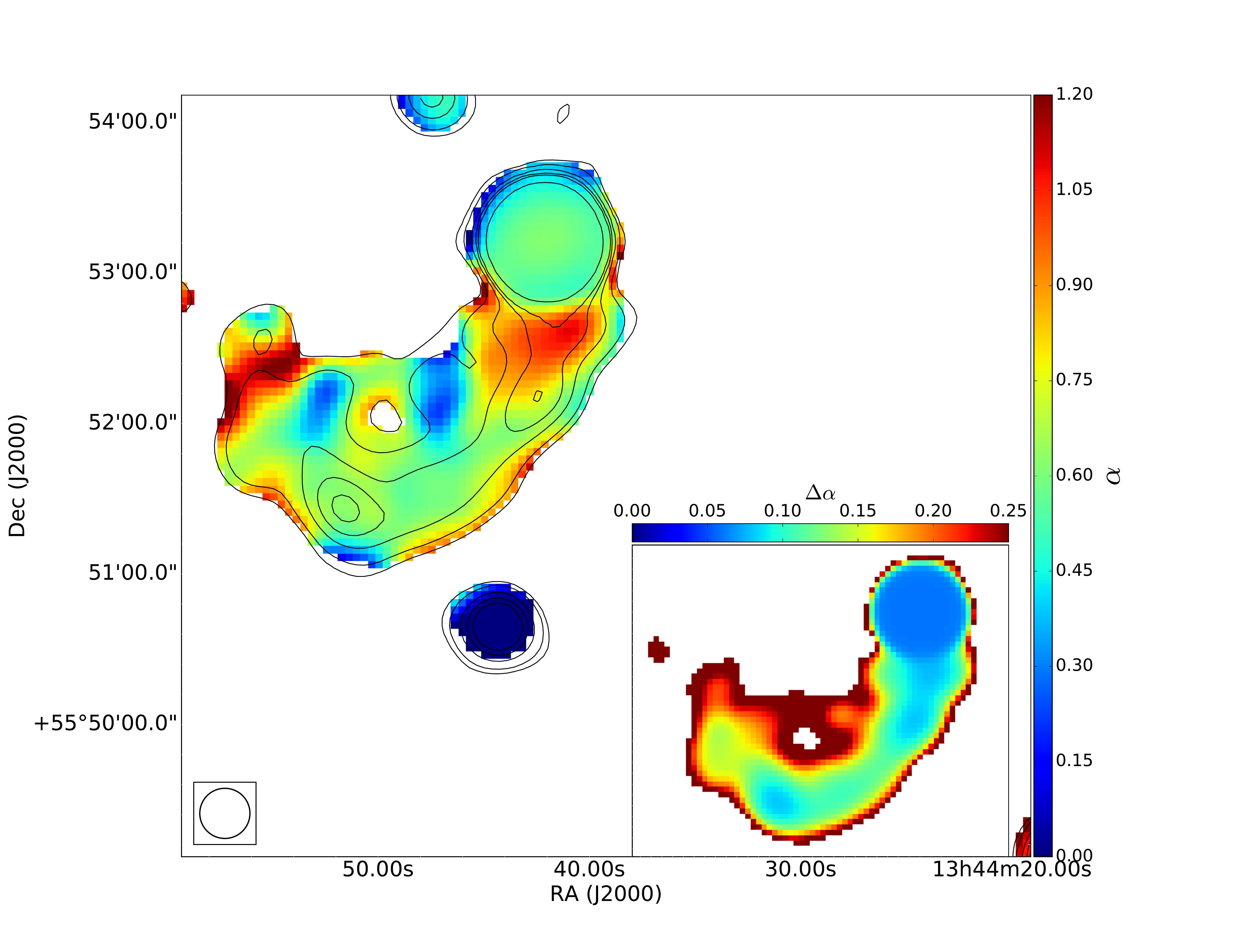}
\caption{20\arcsec~spectral index map from 1455$-$4700\,MHz made with the VLA images. The local RMS noise, $\sigma_{RMS}$, in the 1455\,MHz image is 80 \mujybeam and in the 4700\,MHz image is 40 \mujybeam. The contours mark the 4700\,MHz emission at levels (3, 5, 10, 15, 20,50)$\times\sigma_{RMS}$. The inset map shows the spectral index error.}
\label{spix_4700}%
\end{figure}
Our multi-frequency images allow us to probe the spectral properties of the radio emission from Mrk\,273. Below, we first discuss the spectral properties of the nuclear components and then the large-scale spectral properties. 
\subsubsection{Nuclear region}
 \begin{table*}[h]
   \centering
  \begin{threeparttable}
       \caption[]{Flux densities and spectral indices for nuclear components.}
         \label{components}
     \begin{tabular}{cccccccc}
            \hline
            \hline
            \noalign{\smallskip}
            
            Component & $S_\mathrm{144\,MHz}$  & $S_\mathrm{1400\,MHz}$ & $S_\mathrm{4700\,MHz}$ & $S_\mathrm{8415\,MHz}$ & $\alpha^\mathrm{144}_\mathrm{1400}$& $\alpha^\mathrm{1400}_\mathrm{4700}$ & $\alpha^\mathrm{4700}_\mathrm{8415}$ \\
             & (mJy) & (mJy) & (mJy) & (mJy) & & &\\ 
            \noalign{\smallskip}
            \hline
            \hline
            \noalign{\smallskip}
    \vspace*{5px}
    N  & 50.3$\pm$5.0 & 91.0$\pm$4.6 & 51.6$\pm$2.6 & 36.4$\pm$1.8 & -0.26 & 0.47 & 0.60 \\
    \vspace*{5px}
    SE  & 138.5$\pm$13.9 & 25.0$\pm$1.3 & 7.2$\pm$0.4 & 3.7$\pm$0.2 & 0.75 & 1.03 & 1.14\\
    \vspace*{5px}
    SW  & 25.8$\pm$2.6 & $\geq$~2.0$\pm$0.1 & 1.03$\pm$0.05 & 0.60$\pm$0.03 & $\leq$1.12 & $\geq$0.55 & 0.93 \\
    \vspace*{5px}
    NE  & 32.1$\pm$3.2 & - & 1.9$\pm$0.1 & 1.01$\pm$0.05 & - & - &
    1.08\\
            \noalign{\smallskip}
            \hline
            \hline
     \end{tabular}      
    
      \small{\textbf{Note.} Component column lists the different components as shown in Fig.~\ref{lofar_hba}. The next four columns list the integrated flux densities. The 1400\,MHz flux densities were taken from \citet{Cole1999}. The last three columns list the spectral indices. The 1\,$\sigma$ errors in the spectral indices are $\sim$0.05, $\sim$0.06 and $\sim$0.14 for  $\alpha^\mathrm{144}_\mathrm{1400}$, $\alpha^\mathrm{1400}_\mathrm{4700}$ and $\alpha^\mathrm{4700}_\mathrm{8415}$ respectively.}
     \end{threeparttable} 
   \end{table*}

\label{spec-nuclear}
To study the  spectrum of the components in the nucleus of Mrk\,273 (right panel in Fig.~\ref{lofar_hba}), we estimated their integrated flux densities and calculated their spectral indices. The flux densities for these components were measured from the sub-arcsecond images. For the 1400\,MHz flux density, we used the measurements from \citet{Cole1999}, who studied the 1400\,MHz emission from the nuclear components with MERLIN at $\sim$0.2\arcsec~resolution. We assumed a flux scale error of 5\% on their values.\par

The spectral indices of nuclear components are summarised in Table~\ref{components}. In the N component, we find an inverted spectrum from 144$-$1400\,MHz with $\alpha^\mathrm{144}_\mathrm{1400}$ = -0.26$\pm$0.05, which steepens to $\alpha^\mathrm{1400}_\mathrm{4700}$ = 0.47$\pm$0.06 and $\alpha^\mathrm{4700}_\mathrm{8415}$ = 0.60$\pm$0.14. This tells us that the emission (at low frequencies) is absorbed. The SE component shows a spectral index of $\alpha^\mathrm{144}_\mathrm{1400}$ = 0.75$\pm$0.05, which steepens at higher frequencies to $\alpha^\mathrm{4700}_\mathrm{8415}$ = 1.14$\pm$0.14. This steep spectral index suggests the presence of a radio AGN in this component. This would be in agreement with \citet{Vardoulaki2015}, who constructed a low resolution spectral index map of the nuclear region and proposed the presence of a radio AGN in it. In the SW and NE components, we detect steep spectral indices over all frequency range. The SW component spectrum shows a straight steep spectrum over our frequency range, with $\alpha^\mathrm{144}_\mathrm{4700}$ = 0.92$\pm$0.03 and 
$\alpha^\mathrm{4700}_\mathrm{8415}$ = 0.93$\pm$0.14. The NE component has spectral indices of $\alpha^\mathrm{144}_\mathrm{4700}$ = 0.81$\pm$0.03 and 
$\alpha^\mathrm{4700}_\mathrm{8415}$ = 1.08$\pm$0.14. The NW component is detected in the 144\,MHz and 4700\,MHz image; however, it is not detected in the 1400\,MHz image of \citet{Cole1999} and very faintly detected in the 8415\,MHz image. We estimate a spectral index of $\alpha^\mathrm{144}_\mathrm{4700}$ = 0.92$\pm$0.04 for this component.

\subsubsection{Large-scale emission}
\label{large scale emission}

\textbf{Southern arc}: In the integrated spectral index of the southern arc emission (excluding the nuclear region), we find no significant steepening of the spectrum up to 4700\,MHz, as $\alpha^{54}_{144}$ = 0.71$\pm$0.18, $\alpha^{144}_{1455}$ = 0.74$\pm$0.05 and $\alpha^{1455}_{4700}$ = 0.77$\pm$0.06 (see Fig.~\ref{agemodel} for spectral profiles at different locations in the southern arc). A relative steepening at the high frequency end would be expected in a radiatively ageing population, provided sufficient time has passed. To spatially resolve the spectral properties, we have constructed spectral index maps shown in Fig.~\ref{spix} and~\ref{spix_4700} from 144$-$4700\,MHz. The maps show several interesting features. The first property to note is that while no spectral gradient is observed along the arc, a gradient across the arc can be seen in the $\alpha^{144}_{1455}$ map in Fig.~\ref{spix_six} - from 0.8$-$0.9 ($\pm$0.13) at the inner edge of the arc to 0.5$-$0.6($\pm$0.12) at the outer edge. The spectral index map made with LOFAR 144\,MHz and Apertif 1360\,MHz image at $\sim$16\arcsec~resolution also showed a similar gradient, thus increasing our confidence in the presence of this feature. In the $\alpha^{1455}_{4700}$ map, we detect a steep spectral index of $\sim$1.1$\pm$0.2 at the tail end of the arc. The spectral index from 1455$-$4700\,MHz at the tail end of the southern arc is steeper than the middle region, thus we observe a gradient along the arc at this frequency range. The second property to note is the spectral index of the ridge $\alpha^{144}_{1455}\approx$ 0.81$\pm$0.07, which is steeper than the typical spectral index measured in the arc. The average spectral index for the ridge in the $\alpha^{1455}_{4700}$ map is 0.96$\pm$0.10. In these spectral index maps, the nuclear region shows very little curvature ,with an average spectral index of $\alpha^{144}_{1455}\approx$ 0.45$\pm$0.06 and $\alpha^{1455}_{4700}\approx$ 0.55$\pm$0.08. \par

\textbf{Northern arc}: The sensitivity of the Apertif image allowed us to trace the spectral indices in the northern arc region. As can be seen in Fig.~\ref{spix_twenty}, the northern arc has a typical spectral index of $\alpha^{144}_{1360}\approx$ 0.87$\pm$0.16, within the $3\sigma$ 1360\,MHz contours. This could also have contribution from the background source marked in the figure. Outside the contours, $\alpha^{144}_{1360}\geq$ 0.7$\pm$0.3. Overall, this suggests that the spectral index of the northern arc is likely steeper than its southern counterpart. We do not recover any emission in the northern arc in the 4700\,MHz image, and do not discuss its properties further. We discuss the spectral properties of Mrk\,273 further in Sec~\ref{discussion}.

\subsection{Magnetic field and spectral age}
\label{specage}
 \begin{table*}[h]
   \centering
  \begin{threeparttable}
       \caption[]{Flux densities and model fit results for southern arc and ridge regions}
         \label{largescale}
     \begin{tabular}{ccccccccccc}
            \hline
            \hline
            \noalign{\smallskip}
            
            Region & $S_\mathrm{54\,MHz}$  & $S_\mathrm{144\,MHz}$ & $S_\mathrm{1455\,MHz}$ & $S_\mathrm{4700\,MHz}$  & $\alpha_\mathrm{inj}$ & $B_\mathrm{eq,0}$ & $B_\mathrm{eq,10}$ & $t_\mathrm{spec,0}$ & $t_\mathrm{spec,10}$ & $\chi^{2}_\mathrm{reduced}$ \\
             & (mJy) & (mJy) & (mJy) & (mJy) & & ($\mu$G) & ($\mu$G)  & (Myr) & (Myr) \\ 
            \noalign{\smallskip}
            \hline
            \hline
            \noalign{\smallskip}
    \vspace*{5px}
    R  & 44.4$\pm$6.8 &23.5$\pm$2.3 &3.5$\pm$0.2 & 1.2$\pm$0.07 & 0.65 & 3.5 & 6.7 & 32.1$^{+4.1}_{-2.4}$ & 19.1$^{+2.1}_{-1.7}$ & 0.94\\             
    \vspace*{5px}
    S1  & 20.7$\pm$3.5 &10.3$\pm$1.0 &2.4$\pm$0.1 & 1.2$\pm$0.1 & 0.55 & 1.9 & 3.6 & 22.1$^{+5.5}_{-6.2}$ & 20.0$^{+3.9}_{-6.3}$ & 0.97\\
    \vspace*{5px}
    S2  & 19.5$\pm$3.3 &7.6$\pm$0.8 &1.4$\pm$0.1 & 0.5$\pm$0.05 & 0.61 & 1.9 & 3.7 & 42.0$^{+4.7}_{-5.2}$ & 36.0$^{+3.3}_{-5.1}$ & 1.06\\
            \noalign{\smallskip}
            \hline
            \hline
     \end{tabular}      
    
      \small{\textbf{Note.} Region column lists the different regions as shown in Fig.~\ref{test}. The first four columns list the flux densities. $\alpha_\mathrm{inj}$ column lists the best fit injection index. $B_\mathrm{eq,0}$ and $B_\mathrm{eq,10}$ column list the magnetic field estimate with $\kappa=0$ and $\kappa=10$, respectively. $t_\mathrm{spec,0}$ and $t_\mathrm{spec,10}$ list the spectral ages estimated using $B_\mathrm{eq,0}$ and $B_\mathrm{eq,10}$ respectively. $\chi^{2}_\mathrm{reduced}$ column lists the reduced chi-squared (for two degrees of freedom) for the JP model fit.}
     \end{threeparttable} 
   \end{table*}
   
\begin{figure}
\includegraphics[width=\columnwidth]{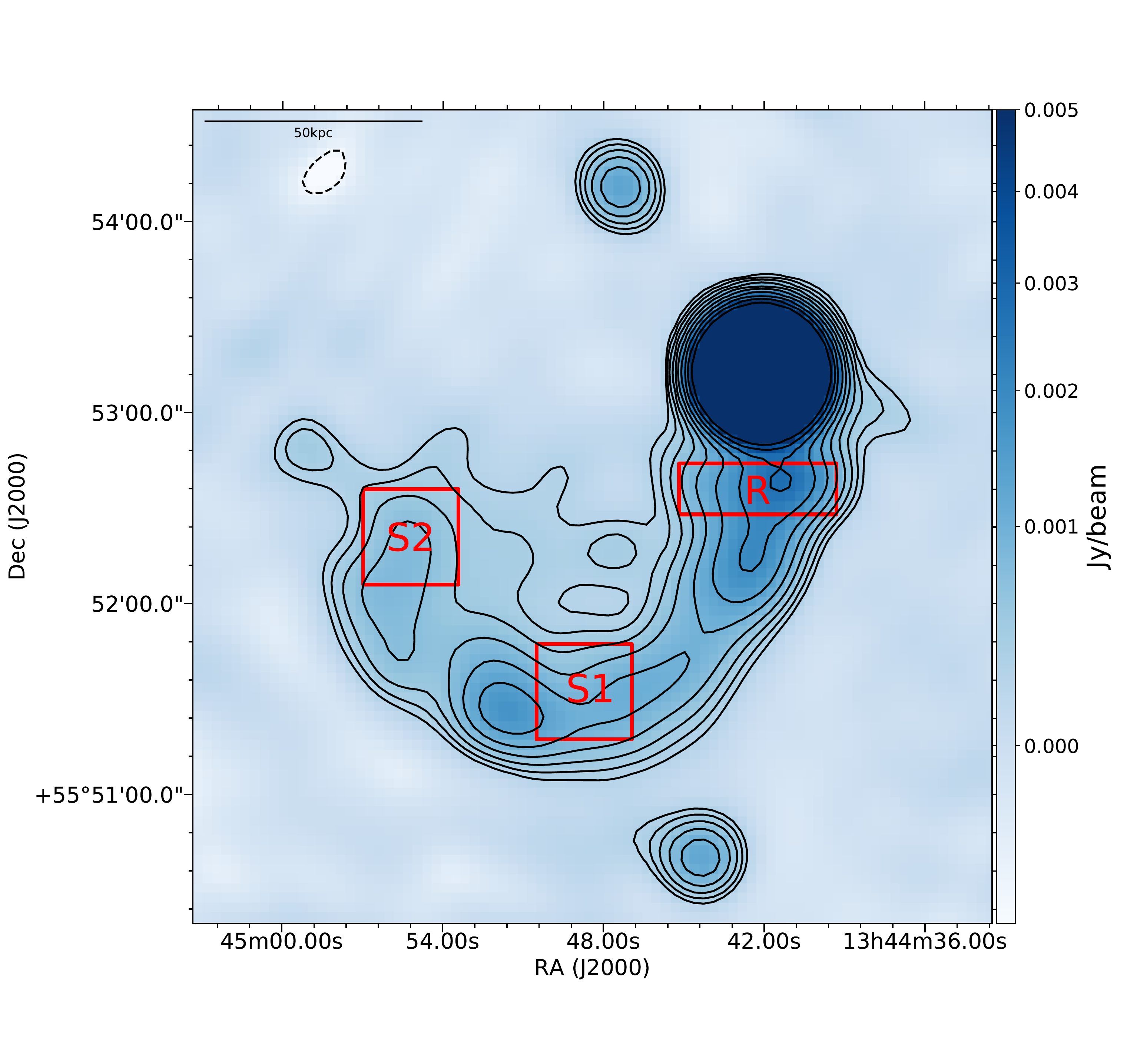}
\caption{VLA 1455\,MHz image of Mrk\,273 at a resolution of 20\arcsec$\times$20\arcsec. The solid lines mark the positive contour levels at: 3$\sigma_\mathrm{RMS}\times\sqrt{2}^{n}$ where n=0,1,2...10 and $\sigma_\mathrm{RMS}$ = 80 \mujybeam. The negative contours are marked by dashed lines and are at $-3\sigma_\mathrm{RMS}$ level. The regions used to extract flux densities in Sec~\ref{specage} are marked and summarised in Table~\ref{largescale}.}
\label{test}
\end{figure}
To obtain the spectral age of the large-scale radio emission in Mrk\,273, we need to first estimate the magnetic field strength. We have used the python version of the SYNCH code \footnote{\url{https://github.com/mhardcastle/pysynch}} \citep{Hardcastle1997} to estimate the equipartition magnetic field in different regions in the source. The regions are marked in Fig.~\ref{test}. We assume that star formation has no contribution to the radio emission in these regions. Typically, while estimating the equipartition magnetic field, the ratio of energy in non-radiating (protons) to radiating particles is assumed to be zero ($\kappa$=0), that is, no non-radiating particles are present. However, studies have shown that in more turbulent lobes of low power radio galaxies, with radio luminosities comparable to Mrk\,273, internal lobe pressure is not sufficient to balance the pressure of the external medium \citep{Croston2003,Dunn2004,Croston2014}. The extra pressure needed is provided by either the higher magnetic field or the entrained material, that is, the non-radiating particles. \citet{Croston2018} found that a ratio of $\sim$10 between the non-radiating to radiating particles was required for the lobe pressure to balance the external pressure. \citet{Bicknell1998} also found that the energy density of radio jets in low power AGNs (like Seyferts) was dominated by thermal plasma. This is likely to be the case for the radio emission in Mrk\,273, given its low luminosity (L$_\mathrm{{1.4\,GHz}}\approx5\times10^{23}$\whz{}) and Seyfert 2 classification. Therefore, we have estimated the equipartition values, reported in Table~\ref{largescale}, using both $\kappa=0$ and $\kappa=10$. Our estimates are in agreement with magnetic field strengths found in similar power radio galaxies \citep{Heesen2018,Das2021}. \par

We have extracted the flux density from all the regions, and used an aged synchrotron spectrum that provided a reasonable fit to the flux densities. For this purpose, we have used all the images at 20\arcsec~resolution. We assumed a cylindrical geometry for all the regions, shown in Fig.~\ref{test}. 
For regions S1 and S2 in the southern arc, we used a length of 30\arcsec~and radius of 15\arcsec. For the region in the ridge, R, we used a length of 50\arcsec~and radius of 8\arcsec. The injection index was varied over the typical range of 0.5$-$0.7, with steps of 0.01, using the Broadband Radio Astronomy Tools (BRATS\footnote{\url{http://www.askanastronomer.co.uk/brats/}}; \citealt{Harwood2013,Harwood2015}) package
to find the best fit injection index ($\alpha_\mathrm{inj}$) for the integrated spectra of the regions. The best fit values for $\alpha_\mathrm{inj}$ were 0.55 for S1, 0.61 for S2 and 0.65 for R. These indices were then used to estimate the magnetic field strengths. These values are listed in Table~\ref{largescale}.
 
\begin{figure}
\includegraphics[width=\columnwidth]{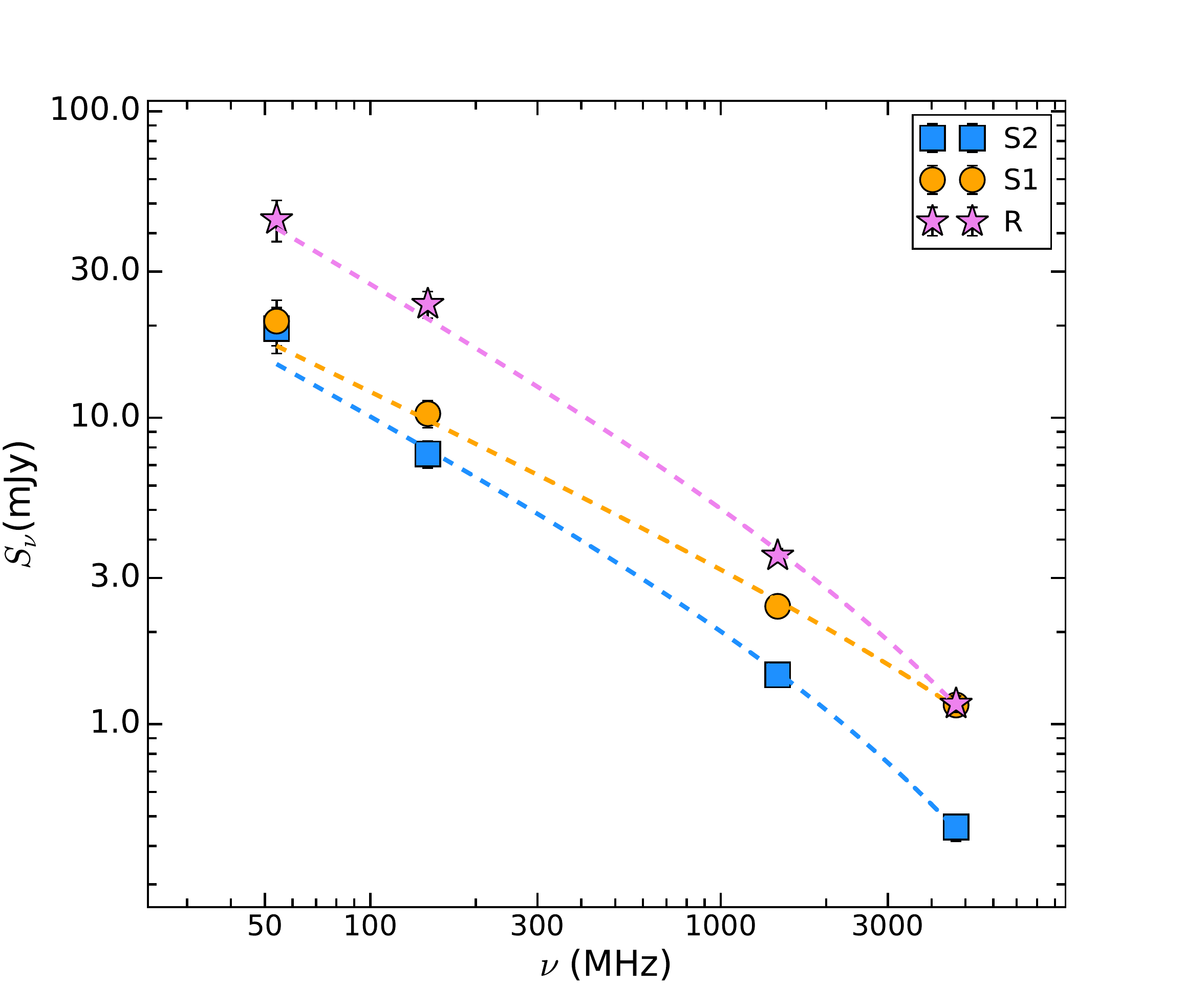}
\caption{JP model fits to the spectrum of the regions marked in Fig.~\ref{test}.}
\label{agemodel}%
\end{figure}

The spectrum of particles emitting synchrotron radiation steepens due to preferential cooling of higher energy particles \citep{Kellermann1966,Pacholczyk1970}. Modelling the steepening of the radio spectra can therefore give us an estimate of the spectral age. If a break exists in the spectra shown in Fig.~\ref{agemodel}, it is above $\sim$5\,GHz. To model these spectra and estimate the spectral ages, we fitted the JP model \citep{Jaffe1973a} to the integrated spectra, using the task \texttt{fitintegrated} in BRATS. We note that there are discrepancies in spectral age between estimates using integrated spectra and a spatially resolved analysis, but our results should still be reliable within the regions. However, these estimates may change for an analysis at lower spatial resolution \citep{Harwood2017}. The model fit results are summarised in Table~\ref{largescale} and shown in Fig.~\ref{agemodel}. We find ages of a few tens of Myr in all the regions. Despite the steeper spectra of the ridge compared to the southern arc, the ages are similar for both due to the higher magnetic field in the ridge. A gradient in spectral age can also be seen from S2 to S1, which is discussed further in Sec~\ref{southern arc}.  \par

In the lobes of radio galaxies, the equipartition magnetic field strength is related to the minimum energy. This can give us information about the properties of the external medium. In order to estimate these properties (and the possible impact of the emission) we use the relation given as 
\begin{ceqn}
\begin{align}
      B_{\mathrm{eq}} =
          \left(\frac{24\pi}{7}u_{min}\right)^{\frac{1}{2}},
\end{align}
\end{ceqn}
where $u_{min}$ is the minimum energy in erg cm$^{-3}$ and $B_{\mathrm{eq}}$ is in Gauss. The particle energy density, derived using the equation above, can be used to approximate the particle density of the surrounding medium. We estimated the relativistic gas pressure as $P_\mathrm{r}$ = ($\gamma$ - 1)$u_\mathrm{min}$, where $\gamma$ = 4/3 for relativistic gas \citep{Pacholczyk1970}, and assumed that the regions are in pressure balance with their surrounding medium. The gas pressure of the surrounding medium can be approximated as $P_\mathrm{m}$ = $\eta\, \mathrm{kT}$, where $\eta$ is the particle density of the medium \citep{Mack1998}. Using the particle energy density obtained above, we roughly estimate a medium particle density of $\eta_{R} = 1\times10^{-3} \mathrm{cm}^{-3}$ for T $=10^{7}$K for the ridge region. This is comparable to the electron number density estimated for the southern X-ray nebula ($\eta_{e}=3\times10^{-3} \mathrm{cm}^{-3}$) that overlaps with the ridge (Fig.~\ref{mlwave}), by \citetalias{Liu2019}. For the southern arc regions, we estimate a particle density of $\sim3\times10^{-4} \mathrm{cm}^{-3}$, an order of magnitude lower than that found in the ridge. Therefore, if such a surrounding medium is present, the low particle density would explain why it is not detected in the soft X-ray images at their current sensitivity.

\section{Discussion}
\label{discussion}
We have traced a rare case of large-scale ($\sim$300\,kpc) radio continuum emission in a ULIRG. The LOFAR Dutch array images have allowed us to trace spectacular and smoothly curved arcs out to $\sim$100$-$200\,kpc, in the north and south, while on kiloparsec scale a complex structure has been traced by the LOFAR International stations data. Using multi-frequency images, we derived the spectral properties from MHz to GHz frequencies, something not done in the very few cases known so far. These data now allow us to explore the properties of the merger system. We first discuss the radio properties of the nuclear components, and explore the possibility of the presence of a radio AGN in this system. Then we move outwards to the large-scale emission and discuss the different scenarios that could explain its origin, spectral properties and how they connect to the nuclear components. Finally, we discuss what this system tells us about presence of nuclear activity in mergers. 

\subsection{Nuclear components}
The sub-arcsecond images show multiple components in the $\sim$4\,kpc nuclear region of Mrk\,273 (Fig.~\ref{lofar_hba}). The multi-wavelength properties of some of these components have been mentioned before in the introduction. The presence of absorbed X-ray AGNs has been confirmed in the N and SW component (\citealt{Iwasawa2011}, \citetalias{Liu2019}), and an AGN has also been proposed in the SE component in the optical emission line study of \citetalias{RodriguezZaurin2014a}. In the radio, an AGN had also been proposed in the nucleus of Mrk\,273 by \citet{Vardoulaki2015} but it could not be pinned down to one nuclear component. \par

\begin{enumerate}
    \item \textbf{N component:} The radio emission of this component is understood to be dominated by a starburst \citep{Bondi2005}. However, \citet{Bondi2005}, in their VLBI study of the nucleus, could not rule out the possibility that one of the compact components could be the radio counterpart of the AGN. As mentioned in Sec~\ref{spec-nuclear}, we detect broadband radio absorption in the N component spectrum. Although the presence of absorption in the nuclear region had been suggested by \citet{Clemens2010}, they did not observe a turnover in the spectrum and could not pinpoint the location of the absorbed component. We note that these differences with respect to our results could be due to the low resolution (5\arcsec and 13\arcsec) of their images, compared to our sub-arcsecond images of the nucleus.
    Using our low frequency sub-arcsecond image of the nucleus, we have detected the absorption to be in the N component.\par
    \vspace{0.1cm}
    Investigating the absorption mechanism, either SSA (synchrotron self-absorption) or FFA (free-free absorption), can give us more information about this component. SSA requires high brightness temperatures, comparable to the kinetic temperatures of $\sim$10$^{10}$\,K of the relativistic electrons \citep{Broderick1975}. \citet{Condon1991} estimated the brightness temperatures of the nuclear components using 0.3\arcsec~images of Mrk\,273 and concluded that the brightness temperatures ($\sim$6$\times10^{3}$\,K) were too low for SSA to be significant. On the other hand, a turnover in the spectrum is also expected in compact, nuclear starbursts due to optically thick FFA \citep{Condon1991,Clemens2010,Leroy2011,Murphy2013}. We therefore consider FFA to be the dominant absorption mechanism for the N component, as also suggested by \citet{Clemens2010}. We used an FFA model that assumes that the absorption is caused by an external homogeneous ionised gas screen, given as 
    \begin{ceqn}
    \begin{align}
    S_\mathrm{\nu} = a \nu^\mathrm{-\alpha} e^\mathrm{-\tau_\mathrm{\nu}},
    \end{align}
    \end{ceqn}

    where $a$ and $\alpha$ are the amplitude and spectral index of the intrinsic synchrotron spectrum, and $\tau_\mathrm{\nu}$ is the optical depth equal to $(\nu/\nu_\mathrm{p})^\mathrm{-2.1}$, where $\nu_\mathrm{p}$ is the frequency at which the optical depth is unity.
\begin{figure}
\includegraphics[width=\columnwidth,height=0.3\textheight]{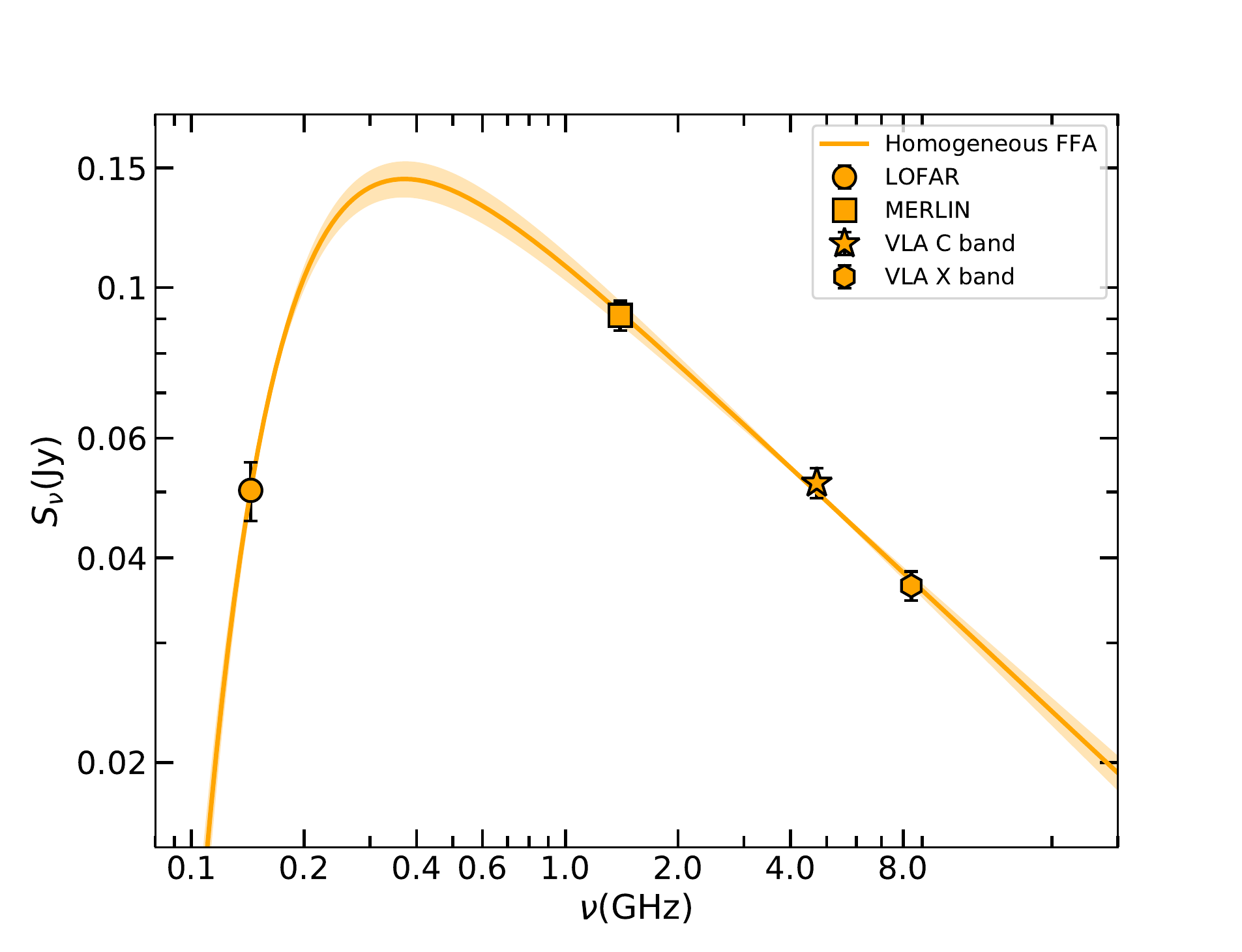}
\caption{Homogeneous FFA model best fit for the N component flux densities. A turnover can be seen at $\sim$370\,MHz.}
\label{abs_N}%
\end{figure}
 We fitted this model to the N component spectrum and found best fit parameters of $a$ = 0.11$\pm$0.01, $\alpha$ = 0.51$\pm$0.03 and $\nu_\mathrm{p}$ = 0.19$\pm$0.01\,GHz. The model fit is shown in Fig.~\ref{abs_N}. For an SSA model fit, we obtained similar model parameters and quality of fit to the data. Using the FFA parameters, we obtained an emission measure of 9.3$\times$10$^{4}$ $\mathrm{pc}\,\mathrm{cm}^\mathrm{-6}$ \citep{ODea1998b}, $\sim$30 times lower than \citet{Clemens2010}. This is due to their assumption of a turnover at $\sim$1\,GHz whereas we estimate the turnover to be at $\sim$370\,MHz. \par
 \vspace{0.1cm}
 Using the deconvolved size from \citet{Condon1991} as the path length for the N component, we estimate an electron density of $\sim$25\,$\mathrm{cm}^\mathrm{-3}$ for the ionised absorbing medium, which corresponds to a column density of N$_{\mathrm{e}}\approx5\times$10$^{21}$ cm$^{-2}$. This is only a fraction of the column density of neutral hydrogen of N$_{\mathrm{H}}=1.7\times$10$^{22}$ cm$^{-2}$ \citep{Cole1999} distributed in a $\sim$250\,pc disc around the starburst. The presence of absorption is interesting because \citet{Nandi2021} found absorption and low spectral ages in several ULIRGs, concluding that these are GPS/CSS sources. However, our analysis shows that in the nuclear component of a ULIRG, where the radio emission is dominated by a starburst, the absorption can also be explained by FFA from the surrounding medium.
\vspace{0.1cm}
\item \textbf{SE component:} The SE component shows a "normal" spectral index at low frequencies, $\alpha^\mathrm{144}_\mathrm{1400}\sim 0.75\pm0.05$, which steepens to $\alpha^\mathrm{4700}_\mathrm{8415}\sim 1.14\pm0.14$. Spectral index of star forming components are not expected to be steeper than $\sim$0.83$\pm$0.13 \citep{Niklas1997}. In their study of spectral properties of ULIRGs, \citet{Vardoulaki2015} proposed the presence of a radio AGN in the nucleus of Mrk\,273 on the basis of a spectral index steeper than that expected from star formation, from 1.4$-$8.4\,GHz. The steep spectral index of the SE component leads us to conclude that the radio AGN proposed by \citet{Vardoulaki2015} is associated with this component. Its radio luminosity of L(144\,MHz)~=~4$\times10^{23}$\whz would make it a low power radio AGN. The idea that this component is a separate nucleus with its own AGN was also suggested by \citetalias{RodriguezZaurin2014a} on the basis of the high optical ionisation state and compact continuum emission. The diffuse emission extending southwards from this component (see Fig.~\ref{lofar_highres}) could be the connection to the southern large-scale features i.e. the arc or the ridge. 
\vspace{0.1cm}
\item \textbf{SW and NE component:} The LOFAR 144\,MHz sub-arcsecond image shows for the first time, resolved structure in the SW component, in the form of two bright spots (see Fig.~\ref{lofar_hba}). Some diffuse emission can be seen around this component in the lower resolution image in Fig.~\ref{lofar_highres}. This component is aligned with the more diffuse emission in the NE component. The SW component is also aligned in the direction of the western outflow, detected by \citetalias{RodriguezZaurin2014a} (see their Figure 1) and \citet{Leung2021a}. Both of these components show steep spectral indices varying from $\sim$0.8$-$1.0. The alignment of these two components and the overlap with optical outflows suggests that this emission could represent outflowing material. The outflows could carry the radio emitting electrons from the nuclear components, or could cause shock-acceleration of the electrons present in the nuclear region. At sub-arcsecond resolution with LOFAR, such steep-spectrum diffuse emission from radio outflows has been detected before in multiple objects, for example Arp\,299 \citep{Ramirez-Olivencia2018,Ramirez-Olivencia2021}, Arp\,220 \citep{Varenius2016} and M82 \citep{Varenius2015}. 
\end{enumerate}

Overall, we propose that the radio AGN in Mrk\,273 is located in the SE component. This adds further evidence for the presence of three AGNs in the nucleus of Mrk\,273. The low resolution image in Fig.~\ref{lofar_highres} shows diffuse emission around the nuclear components that had been missed by the previous sub-arcsecond images at GHz frequencies. The southward extension from the SE component, could be the connection between the radio AGN in the SE component and the southern arc or ridge.

\subsection{Large-scale}
\label{large scale}
The large-scale structure can be described as a combination of three components - the arcs in north and south, and the ridge. The direction of the southern arc, partly consistent with the optical tail at that location (see Fig.~\ref{mlwave}), and perpendicular to the ridge, suggests that the southern arc and the ridge are not physically related. \citetalias{Liu2019} also proposed that the soft X-ray nebula, that overlaps with the ridge, is unrelated to the tidal tail that casts a shadow over the nebula emission. Therefore, we discuss these separately below. 

\subsubsection{Southern arc}
\label{southern arc}
Starbursts and AGNs are known to be the dominant power source in ULIRGs. In Mrk\,273, the high SFR and the presence of at least two to possibly three AGNs in the nuclear region makes it necessary to explore both as the powering mechanisms of the radio emission. Below, we discuss the various scenarios for the origin of the southern arc emission.\par

\textbf{Star formation:} Radio emission from galaxies also has contribution from star formation, which emits both thermal (free-free) and non-thermal (synchrotron) radiation. The thermal component dominates above $\sim$30\,GHz, and has a spectral index of $\sim$0.1. The non-thermal emission, from relativistic electrons accelerated by supernova remnants, dominates the spectrum up to $\sim$10\,GHz. Assuming that all the radio emission in the arc at 1.4\,GHz is from star formation, and using the calibration from \citet{Yun2001}, we estimate an SFR of 24.2$\pm$7.4 M\textsubscript{\(\odot\)} yr$^{-1}$. Although such high SFRs are seen in some ULIRGs, they are not expected at distances of $\sim$100\,kpc from the nucleus, in the form of an arc. This SFR corresponds to an H$\alpha$ surface brightness of $\sim$6$\times10^{-17}$ erg s$^{-1}$ cm$^{-2}$ arcsec$^{-2}$ \citep{KennicuttJr.1998}. \citetalias{Spence2016} study the H$\alpha$ emission in Mrk\,273, and have a sensitivity of $\sim$1$\times10^{-17}$ erg s$^{-1}$ cm$^{-2}$ but recover no H$\alpha$ emission coinciding with the southern arc, out to 100\,kpc. There is also no optical continuum emission detected along the arc from the stars being formed. No emission from dust has been detected along the arc either, which could hide the optical emission. The optical continuum and H$\alpha$ emission is only seen out to a radial extent of $\sim$45\,kpc to the south of the nucleus \citepalias{Spence2016}, which overlaps with the radio ridge. This tells us that star formation in the arc is not the origin of the radio emission.\par
\textbf{Tidal origin:} Another possibility is that the southern (and northern) arc have tidal origin. In this scenario, the radio emission would be associated with supernovae from the stars stripped from the merging galaxies. However, a lack of emission from stars (no optical continuum) and gas (no HI or ionised gas emission) over these regions makes it unlikely (\citetalias{RodriguezZaurin2014a,Spence2016}; \citealt{Leung2021a}). The dynamical timescale of the merger is also much larger than the typical  synchrotron lifetime of $\sim$40\,Myr for the emitting electrons. Therefore, if the southern arc was made up of tidally stripped relativistic plasma left behind as the merger progressed, there would be more signs of spectral ageing, which we do not detect. This suggests that the southern arc does not have a tidal origin.\par
\textbf{Galactic winds:} Nuclear starbursts and AGNs in galaxies can also drive large-scale galactic winds, that can be detected at different wavelengths \citep{Heckman1990a,Veilleux2005}. In the last few years, diffuse emission from starburst driven galactic winds has been detected at large radial distances from the nucleus, including at $\sim$60\,kpc \citep{Hodges-Kluck2020} and $\sim$50\,kpc \citep{Rupke2019}, although the actual sizes could be even larger since these observations are limited by their field of view.\par In the case of Mrk\,273, recent wide-field integral field unit observations of optical emission lines have yielded no such detections \citep{Leung2021a} out to a radial extent of $\sim$40\,kpc. Our high sensitivity low frequency radio observations could be more sensitive to such winds. Therefore, the southern arc could be a result of galactic winds originating from the nucleus, ploughing into the external medium. The curvature of the southern arc could be caused by the interaction of these winds with an external medium. We refer to this as Scenario A.  Given the presence of powerful AGNs in the nucleus, we propose that these winds could be driven by AGNs, but some contribution from the nuclear starburst cannot be ruled out. However, compared to known cases of large-scale galactic winds, the southern arc has a much more collimated morphology and a significantly larger total spatial extent of $\sim$300\,kpc (southern + northern arc). Although we cannot rule out this scenario, the morphological properties reduce our confidence in it.\par

\textbf{AGN:} The more likely case is that the southern arc is fuelled with relativistic electrons by an AGN in Mrk\,273. The luminosity of the southern arc, L(144\,MHz) = 2$\times 10^{23}$\whz, is typical of radio lobes in AGNs, although it lies on the "radio quiet" end of the population \citep{Panessa2019}. \citet{Yun1999} have also suggested that the luminosity and high degree of polarisation of the extended emission in Mrk\,273 points towards an AGN origin. In this case, the southern arc would be fuelled by a radio AGN in the nucleus, i.e. the SE component. However, the radio AGN could have either already been present in one of the merging galaxies $-$ which we refer to as Scenario B $-$ or could have been triggered by the merger $-$ which we refer to as Scenario C. \par
\begin{enumerate}
\item In Scenario B, the smooth curvature of the arc would then trace the infall of the radio AGN. But such plasma left behind as the merger progresses would also be expected to show signs of strong ($\alpha\geq1.2$ or $\Delta\alpha\geq0.5$) spectral ageing, as discussed above, which we do not detect. 
We do find that the spectral index is steeper at the tail end (S2) than at the leading end (S1, Sec~\ref{large scale emission}). The spectral age of the arc emission is also higher, 36$-$42\,Myr, at the tail end (S2) than at the leading end (S1), 20$-$22\,Myr. If we assume that the age difference is due to the motion of the infalling galaxy as mentioned above, and use the difference in the spectral ages as the time taken by the infalling galaxy to cover the distance between the regions, we can approximate, to first order, an infall speed of the galaxy ($v\approx$ $d$/$t_\mathrm{spec}$). Using 55\,kpc for the distance between the two regions, we estimate an infall speed of $\sim$2500$-$3300\,\kms. During interaction of galaxies in non-cluster environments, the velocities are expected to be of the order of the typical gravitational motions in galaxies, i.e. several hundred \kms. For instance, the collisions in Taffy galaxies are understood to occur at $\sim$600$-$800\,\kms (see \citealt{Condon1993,Vollmer2012}). However, the velocity we estimate, albeit a first order approximation, is much higher than seen before for interacting galaxies. Therefore, if scenario B were true, it points to other processes, like in situ particle reacceleration, being present in the arc. Such processes have also been suggested to be present in the nearby ULIRG Mrk\,231.\par
\vspace{2.5mm}
\item In Scenario C, the southern arc would be a radio lobe fuelled by a low power radio jet from AGN activity triggered by the merger, in the pre-coalescence stage. The low power means that the lobe could be more turbulent and less well collimated than in typical radio galaxies with powerful and highly collimated radio jets. A lobe fuelled by a radio jet can explain the large spatial extent, since the energy flux can be channelled better than in the case of winds, allowing the plasma to travel greater distances with relatively less energy loss. A radio jet would also produce a much more collimated morphology than AGN driven winds. The smooth curvature of the southern arc in this case would be due to interaction with the large-scale external medium bending the low power radio lobe. \par 
We note that we do not directly detect the presence of a radio jet in Mrk\,273 in our sub-arcsecond images. This could be due to the jet getting decollimated, as it makes its way out of the nucleus, to a level where its surface brightness is below the sensitivity of our sub-arcsecond images. However, we do see an extension of the SE component towards the southern arc, in the smoothed LOFAR sub-arcsecond image (Fig.~\ref{lofar_highres}), as mentioned in Sect.~\ref{nuclear_morphology}.
\end{enumerate}

A new feature observed in the southern arc is the presence of a spectral gradient across the arc, with $\alpha$ $\sim$ 0.6 at the outer edge and $\sim$0.9 at the inner edge (Fig.~\ref{spix_six}). We also note an edge-brightened feature at the outer edge in the 6\arcsec~contour plot at 144\,MHz. This suggests that the plasma is interacting with an external medium at this location, which rejuvenates the electrons at the outer edge. This would be expected if the plasma were moving radially outwards, like the wind in Scenario A or radio jet in Scenario C, rather than being formed locally as expected in Scenario B. Therefore, our spatially resolved spectral properties help us rule out Scenario B. Assuming that the southern arc plasma is in pressure balance with such a surrounding medium, we estimate the particle density of the medium to be lower (by an order of magnitude) than the southern X-ray nebula \citepalias{Liu2019}. This could be a reason why such a large-scale medium has not been detected in the current soft X-ray images. \par

Using our spectral age estimate at the tail end of the southern arc (S2), we can also approximate, to first order, the speed of the galactic winds required in Scenario A or radio lobe in Scenario C. Assuming constant velocity, a distance of 100\,kpc from the nucleus for S2, and a time range of 36$-$42\,Myr gives a velocity range of $\sim$2300$-$2700 \kms. For galactic winds in Scenario A, such a high speed at a distance of 100\,kpc from the nucleus is highly unlikely. For instance, the speed of galactic wind at a distance of $\sim$50\,kpc in SDSS\,J211824.06+001729.4 (Makani; \citealt{Rupke2019}) is only 120 \kms, and for NGC\,3079 at a distance of $\sim$60\,kpc it is 500 \kms \citep{Hodges-Kluck2020}. However, for radio lobe in Scenario C, our estimate corresponds to 0.5$-$0.7\% of the speed of light. This is in agreement with the current understanding of radio jets, which are understood to decelerate to sub-relativistic speeds at tens of kiloparsecs distance from the nucleus \citep{Laing2002b,Laing2014,Hardcastle2020}. Therefore, we propose Scenario C as a more favourable explanation for the origin of the southern arc. \par

As mentioned before, large-scale radio continuum emission has also been detected in the ULIRG Mrk\,231, also known to host an AGN and a starburst. \citet{Morganti2016} had concluded that the large radio lobes could be fuelled by electrons from the central AGN by or from electrons accelerated in situ. Although \citet{Morganti2016} did not directly detect a radio jet from their data either, they detected a "bridge" of emission connecting the lobe to the nucleus, which supports the fuelling from central AGN hypothesis. Although the radio power of Mrk\,273 is lower than Mrk\,231 (L$_\mathrm{1.4\,GHz}$ = 1.4$\times10^{24}$\whz), the presence of multiple AGNs in the nucleus, including one radio AGN (in the SE component), suggests that processes similar to Mrk\,231 could also be responsible for the radio emission in the southern arc of Mrk\,273.\par

\subsubsection{Northern arc}
\label{northern arc}
We detect, for the first time, a giant $\sim$190\,kpc arc in the north of the nuclear region of Mrk\,273, which is much larger than its southern counterpart ($\sim$100\,kpc). This component is also more diffuse and has lower surface brightness than the southern arc. It also does not show a smooth curvature. A background galaxy 2MFGC\,11079 at $z=0.038$ also contributes to the radio emission in this arc, marked in Fig.~\ref{lofar_hba}. In Scenario A, the northern arc would be formed by the counter-wind as it expands into the external medium. Scenario B cannot explain the origin of the northern arc, since the southern arc is created locally in that scenario. In Scenario C, the northern arc would be formed by the radio lobe fuelled by the counter jet.
\par
In Sec~\ref{large scale emission}, we found that the spectral index in this arc is likely steeper than the southern arc. Given the low power of the radio AGN involved and the less collimated structure of the northern arc, it is unlikely that this is a result of relativistic beaming. Instead, the less collimated structure and larger size of the northern arc could be due to different conditions of the surrounding medium (i.e. halo material, circumgalactic medium) around it, compared to the south. A less dense medium in this region would mean that the wind (Scenario A) or radio lobe (Scenario C) could travel larger distances in the same time as the southern arc. The scarcity of interactions with a dense surrounding medium could cause a lack of rejuvenated electrons responsible for the radio emission, leading to the lower surface brightness and steeper spectral index (since the feature is not detected at 4700\,MHz). The lack of curvature in this arc would also be explained by the lack of a medium in this location to bend the plasma. Such asymmetric radio emission has also been seen before in the ULIRG Mrk\,231 \citep{Morganti2016}. 

\subsubsection{Ridge}
\label{ridge}
A peculiar feature in Mrk\,273 is the ridge of radio emission, at a distance of $\sim$25\,kpc south of the nucleus. This overlaps very well with the soft X-ray emission, shown in Fig.~\ref{mlwave}, and also the ionised gas nebula observed in this region \citepalias{RodriguezZaurin2014a,Spence2016}. The emission in the nebula is understood to be due to shocks in this region. \citetalias{RodriguezZaurin2014a} and \citetalias{Spence2016} propose that the gas in the nebula is ionised by low velocity shocks (200$-$400\,\kms{}) and \citetalias{Liu2019} conclude that the gas is heated to X-ray emitting temperatures by shocks that sweep through the medium. \citetalias{Liu2019} also propose that the X-ray nebula in this region would have to be enriched by multiple outflows. We propose that the radio emission in this region is due to a reservoir of electrons deposited by multiple outflows, and then accelerated by the shocks that sweep the medium. The spectral age of the ridge, $\sim$20$-$32\,Myr, is also in agreement with the timescale of enrichment of the nebula by multiple outflows, $\lesssim$0.1\,Gyr, estimated by \citetalias{Liu2019}. The radio ridge also overlaps with the tidal tail seen in optical continuum, and therefore has some contribution from the star formation in the tail.\par
The systematically steeper spectral index of the ridge in comparison to the arc ($\approx$ 2$-$3$\sigma$ significance) suggests that these components are formed by two separate electron populations. The electron population that makes the ridge has a steeper energy distribution, i.e. less high energy electrons relative to the arc, leading to a steeper spectral index. This could be due to the different powering mechanisms for the ridge and the southern arc. The southern arc is fuelled by a radio AGN, which would be more efficient at accelerating electrons than the shocks present in the ridge. 
\par
We do not observe any radio emission to the north-east of the nucleus, where another ionised gas and soft X-ray nebula is seen (Fig.~\ref{mlwave}). \citetalias{RodriguezZaurin2014a} find quiescent gas kinematics in this region and conclude that the ionised gas represents tidal debris left over from a secondary merger event illuminated by one of the AGNs from the nucleus. \citetalias{Liu2019} also propose that this nebula is not enriched by multiple outflows. The lack of radio emission then, would be in line with these observations, since there would be no pool of electrons deposited by outflows in this region and no shocks to accelerate them to synchrotron emitting energies.

\subsection{Mrk\,273 in the broader context of ULIRGs}

Mrk\,273 shows spectacular continuum and spectral properties on both large and small-scales. Although extended optical continuum and HI emission is known to occur in major mergers, radio continuum from such objects is still very rare (for a review, see \citealt{Hibbard2001}). Similarly, very few cases of extended radio emission have been identified in ULIRGs (\citetalias{Yun1999}; \citealt{Hayashi2021,Nandi2021}), and even fewer have been spatially resolved. To the best of our knowledge, this is the first time the spectral properties of the resolved extended emission from an ULIRG have been studied (similar analysis was done for the LIRG Arp\,299 by \citealt{Ramirez-Olivencia2021}). We propose that the southern arc is a formed by a radio AGN in the nuclear region of Mrk\,273, and a star formation or tidal origin for the emission are unlikely. On the basis of the steep spectral index, we conclude that this radio AGN is associated with the SE component in the nucleus. In their high resolution study of the nuclear components, \citet{Bondi2005} also suggested that the morphology of the SE component resembled a radio AGN. \par

The presence of radio activity in Mrk\,273 can be useful to link radio AGNs and ULIRGs. We find that fuelling from a radio AGN can explain the formation of both the arcs, where the radio activity has been triggered by the merger. This shows that activity can be triggered even before the nuclei have coalesced and strengthens the current understanding that activity can be triggered at different stages of the merging process \citep{Tadhunter2011}. Another link between ULIRGs and radio AGNs is that in the local universe, ULIRGs may be the early stage in the evolution of a merger into a radio galaxy \citep{Nandi2021}. Our results for Mrk\,273 support this hypothesis. Although it does not appear very likely, another possibility is that the arcs are formed by AGN driven high speed winds. However, if this is the case, then this would be one of the largest known cases of galactic winds, with a total spatial extent of $\sim$300\,kpc. In any case, our results, combined with the other ULIRGs mentioned in Sec~\ref{introduction}, give further evidence for the idea that large ($\sim$300\,kpc) radio structures in ULIRGs are signposts of AGNs in the system. 
  \par
To further test this hypothesis, it is important that a large number of ULIRGs are inspected for the presence of large-scale low surface brightness radio emission. For this purpose, observations with high sensitivity to low-surface brightness features are needed. As illustrated by this work, low frequency surveys are particularly suited for this and, in particular, the ongoing LOFAR surveys LoLSS and LoTSS can provide these data. However, our work has also illustrated the importance of the spectral analysis for understanding the origin of the radio structure in ULIRG. \par

In the case of the present work, the addition of the image from the Apertif survey has allowed us to derive the (spatially resolved) spectral properties of the extended emission in Mrk\,273. The subsequent confirmation (and refinement) provided by the pointed VLA observations is an important step. This illustrates the potential of the LOFAR-Apertif combination for studying a large sample of these objects. Other surveys, in particular EMU at $\sim$800\,MHz, will provide comparable images for an even larger sky area. Studying the spectral properties is useful because it can help pin down the origin of the radio emission and provide an estimate for the timescale of merger triggered activity. This information would be crucial for our understanding of the relation between mergers and AGNs.

\begin{acknowledgements}
We would like to thank the referee, Loreto Barcos-Munoz, for their comments, especially on the VLA C band data reduction, which has significantly improved the quality of the paper. LOFAR, the Low Frequency Array designed and constructed by ASTRON, has facilities in several countries, that are owned by various parties (each with their own funding sources), and that are collectively operated by the International LOFAR Telescope (ILT) foundation under a joint scientific policy. This work makes use of data from the Apertif system installed at the Westerbork Synthesis Radio Telescope owned by ASTRON. ASTRON, the Netherlands Institute for Radio Astronomy, is an institute of the Dutch Science Organisation (De Nederlandse Organisatie voor Wetenschappelijk Onderzoek, NWO). This work was supported by the Medical Research Council [MR/T042842/1]. CT acknowledges support from STFC. AD acknowledges support by the BMBF Verbundforschung under the grant 05A20STA. RJvW acknowledges support from the ERC Starting Grant ClusterWeb 804208. FdG acknowledges support from the Deutsche Forschungsgemeinschaft under Germany's Excellence Strategy - EXC 2121 “Quantum Universe” - 390833306. 
EAKA is supported by the WISE research programme, which is financed by the Dutch Research Council (NWO). JvL acknowledges funding from Vici research programme `ARGO' with project number 639.043.815, financed by the Dutch Research Council (NWO). LCO acknowledges funding from the European Research Council under the European Union's Seventh Framework Programme (FP/2007-2013)/ERC Grant Agreement No. 617199. DV acknowledges support from the Netherlands eScience Center (NLeSC) under grant ASDI.15.406. JMvdH acknowledges funding from the Europeaní Research Council under the European Union’s Seventh Framework Programme (FP/2007-2013)/ERC Grant Agreement No. 291531 (‘HIStoryNU’). The J\"ulich LOFAR Long Term Archive and the German LOFAR network are both coordinated and operated by the J\"ulich Supercomputing Centre (JSC), and computing resources on the supercomputer JUWELS at JSC were provided by the Gauss Centre for Supercomputing e.V. (grant CHTB00) through the John von Neumann Institute for Computing (NIC). KMH acknowledges funding from the State Agency for Research of the Spanish Ministry of Science, Innovation and Universities through the ``Center of Excellence Severo Ochoa'' awarded to the Instituto de Astrof\'isica de Andaluc\'ia (SEV-2017-0709); from grant RTI2018-096228-B-C31 (Ministry of Science, Innovation and Universities / State Agency for Research / European Regional Development Funds, European Union); and from the coordination of the participation in SKA-SPAIN, funded by the Ministry of Science and innovation (MICIN).

\end{acknowledgements}

  \bibliographystyle{aa-copy} 
  \bibliography{References}

\begin{appendix}

\section{Continuum images}

\begin{figure*}[t!]
        \centering
        \begin{subfigure}{0.49\textwidth}
            \includegraphics[width=\textwidth,height=0.28\textheight]{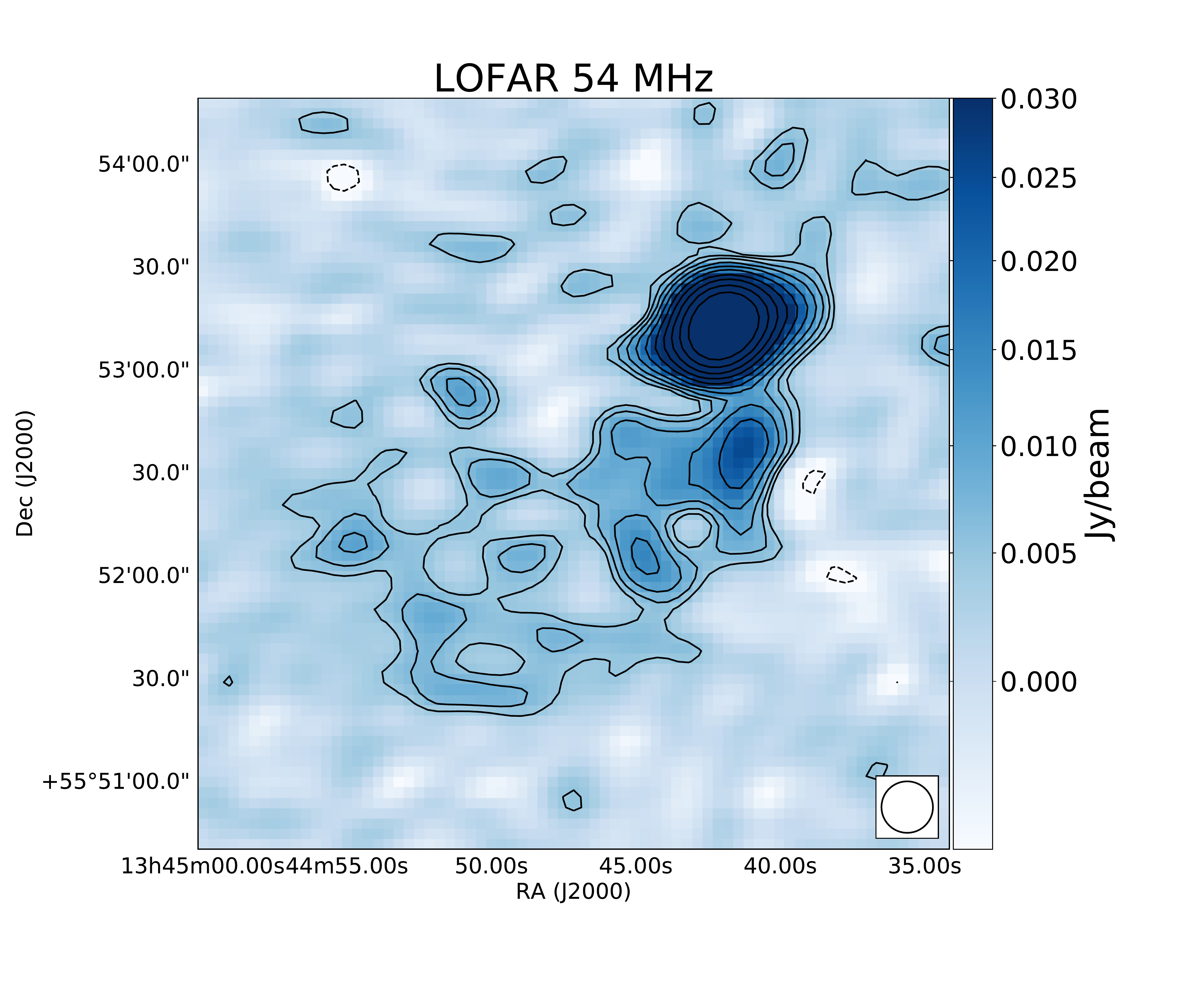}
            \caption[1]%
            {}    
            \label{lba}
        \end{subfigure}
        \centering
        \begin{subfigure}{0.49\textwidth} 
            \includegraphics[width=\textwidth,height = 0.28\textheight]{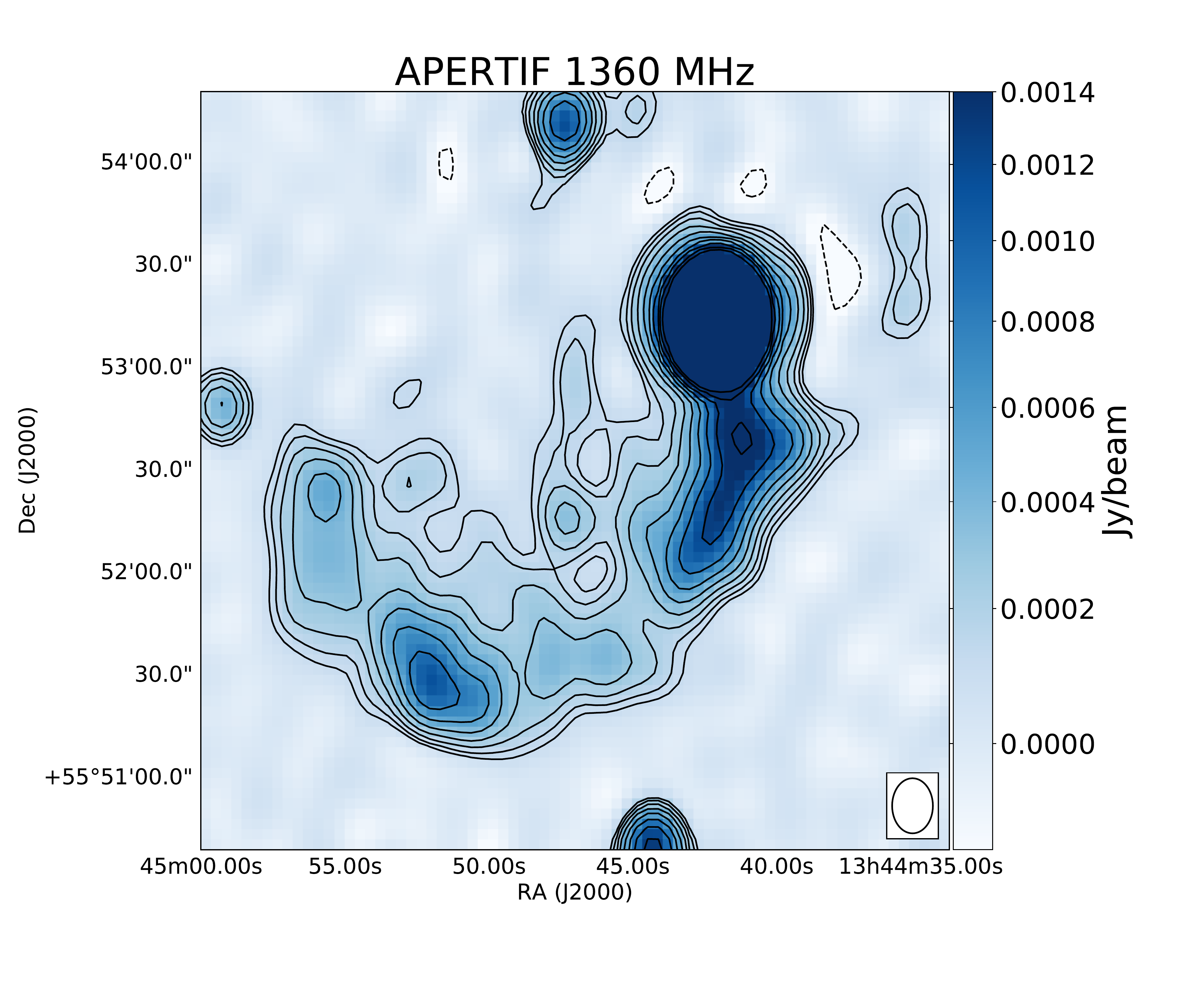}
            \caption[2]%
            {}  
            \label{apertif}
        \end{subfigure}
        \centering
        \begin{subfigure}{0.49\textwidth} 
            \includegraphics[width=\textwidth,height = 0.28\textheight]{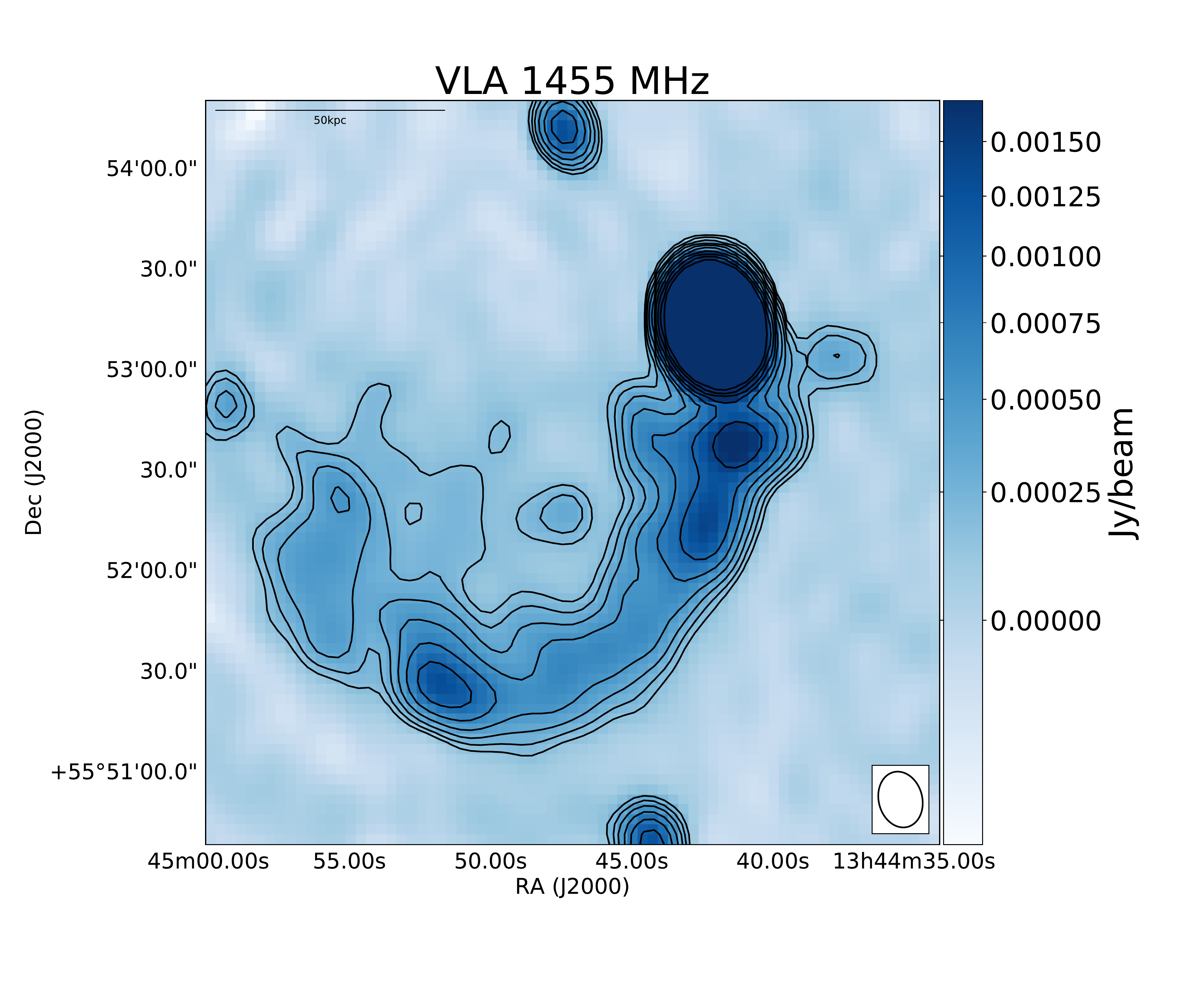}
            \caption[3]%
            {}  
            \label{vla_lband}
        \end{subfigure}     
        \centering
        \begin{subfigure}{0.49\textwidth} 
            \includegraphics[width=\textwidth,height = 0.28\textheight]{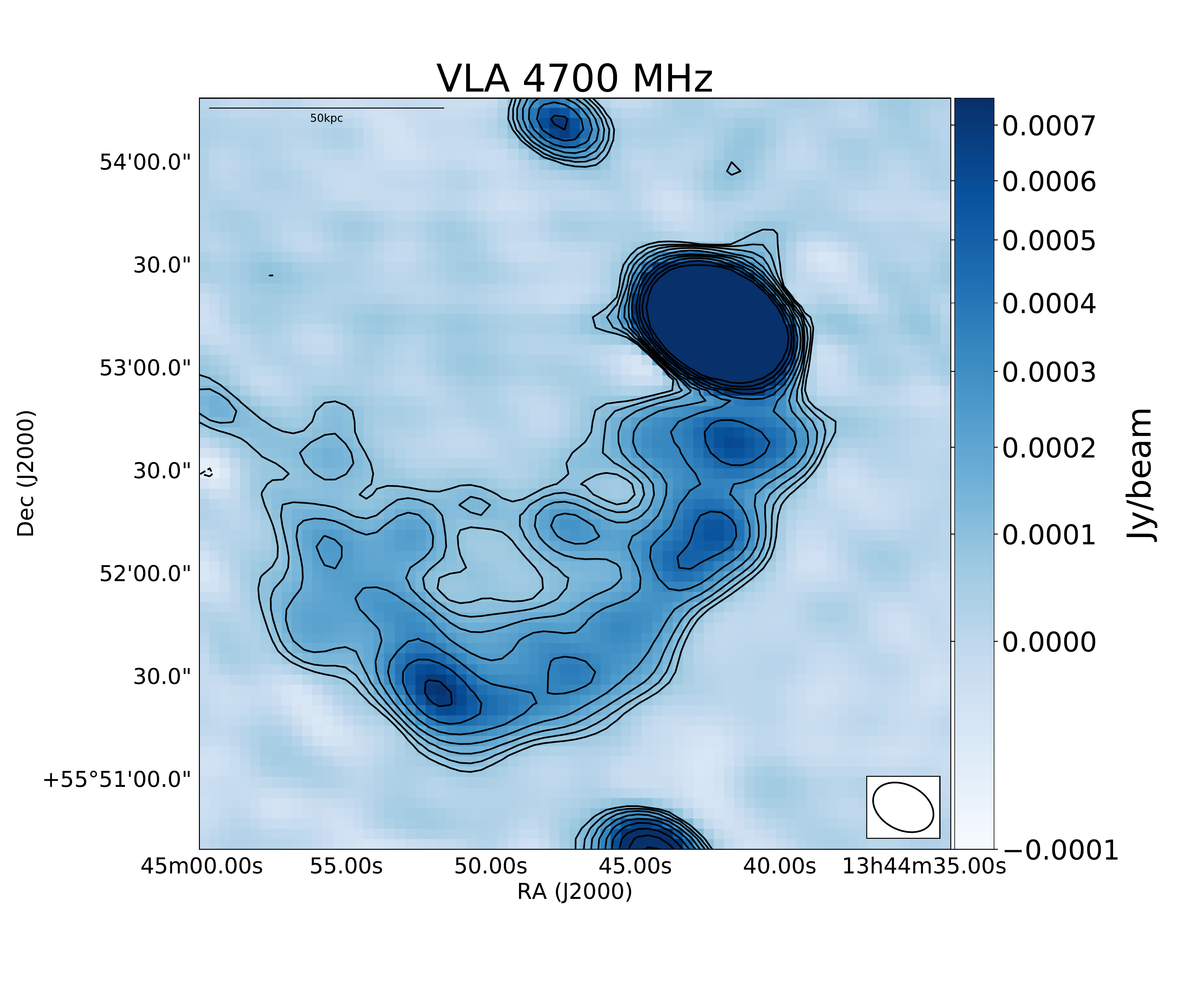}
            \caption[4]%
            {}  
            \label{vla_cband}
        \end{subfigure} 
        \centering
        \begin{subfigure}{0.49\textwidth}
            \includegraphics[width=\textwidth,height=0.28\textheight]{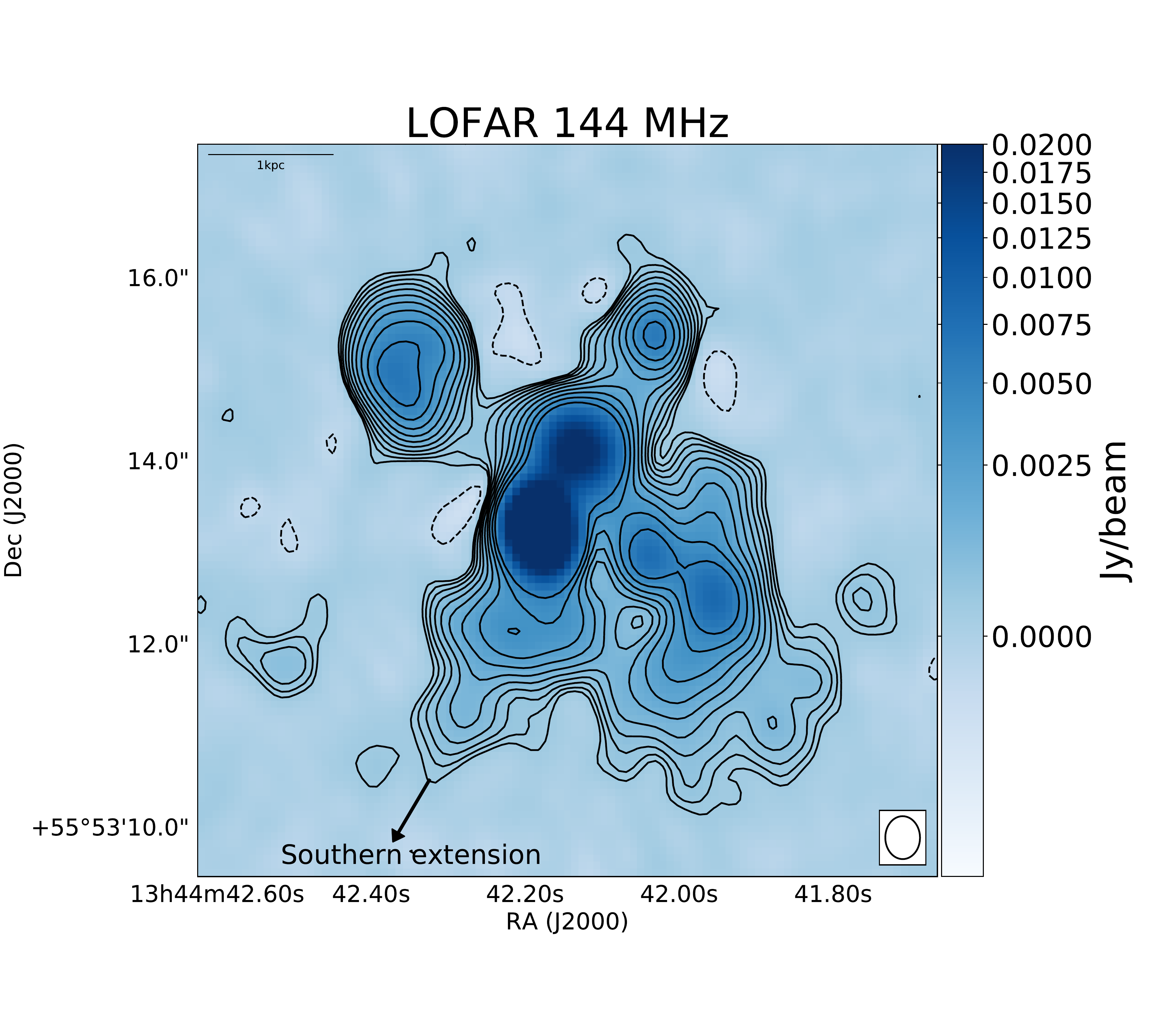}
            \caption[5]%
            {}    
            \label{lofar_highres}
        \end{subfigure}        
        \centering
        \begin{subfigure}{0.49\textwidth}
            \includegraphics[width=\textwidth,height=0.28\textheight]{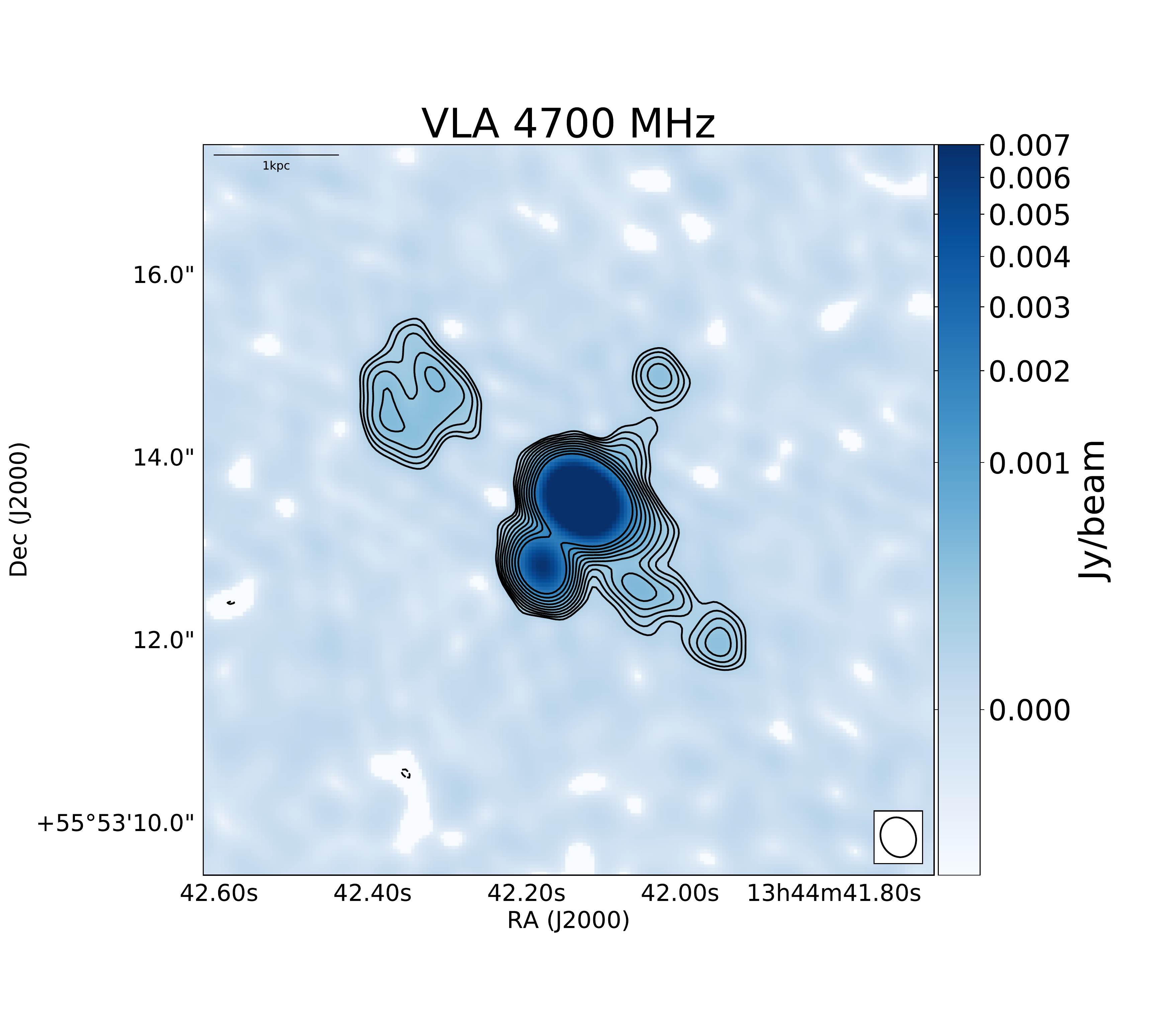}
            \caption[6]%
            {}    
            \label{cband_highres}
        \end{subfigure}
\end{figure*}
\begin{figure*}
\ContinuedFloat
        \centering
        \begin{subfigure}{0.49\textwidth} 
            \includegraphics[width=\textwidth,height = 0.28\textheight]{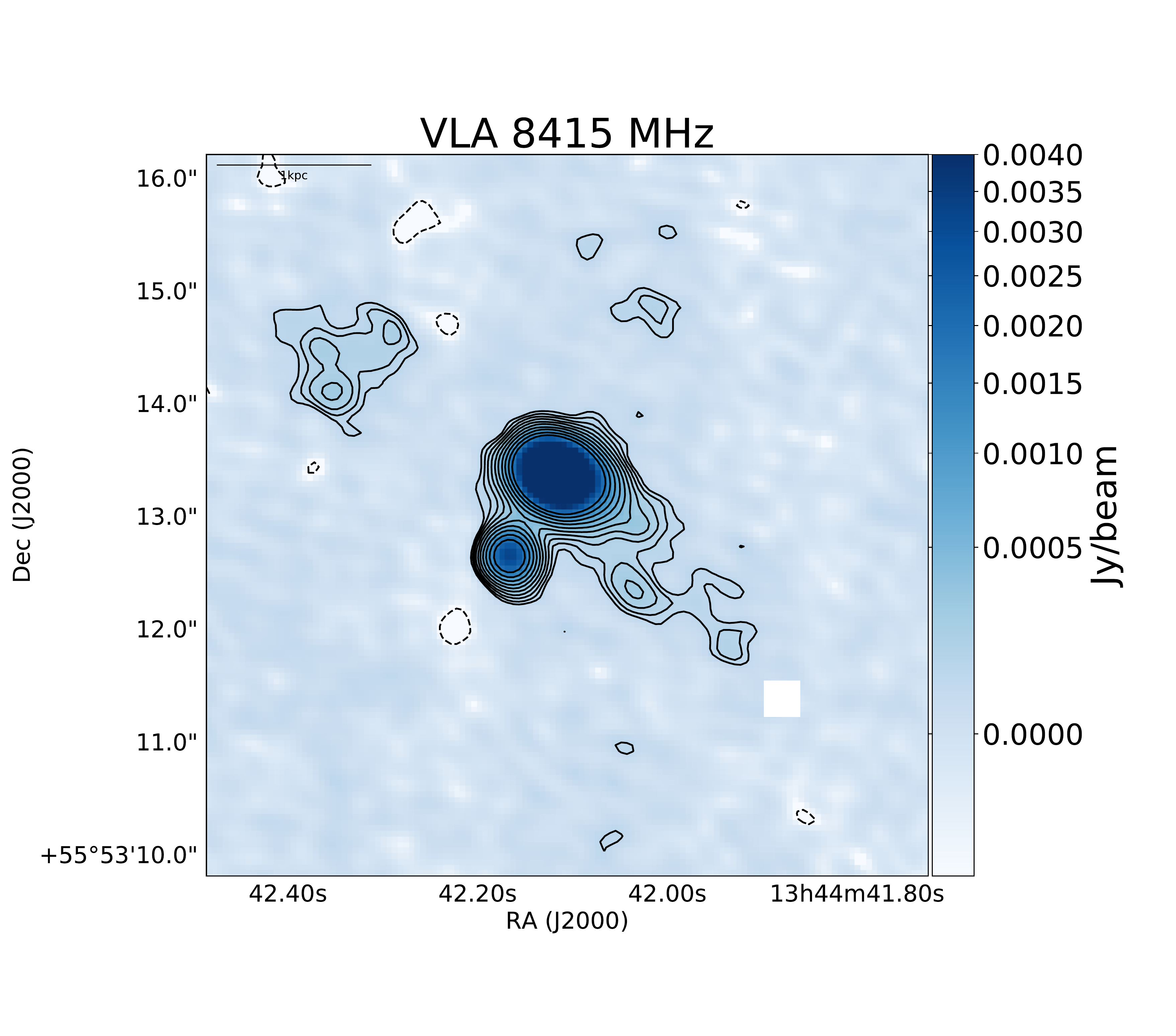}
            \caption[7]%
            {}  
            \label{xband_highres}
        \end{subfigure}
        \caption[international]   
        {\small Continuum images of Mrk\,273. Images in the first and second row show the large-scale structure at (a) 54\,MHz with LOFAR LBA, (b) 1360\,MHz with Apertif, (c) 1455\,MHz with VLA and (d) 4700\,MHz with VLA. The third and fourth row show small-scale structure in the nuclear region with (e) LOFAR at 144\,MHz, VLA at (f) 4700\,MHz and (g) 8415\,MHz. The contour levels in all images are 3$\sigma_\mathrm{RMS}\times\sqrt{2}^{n}$ where n=0,1,2...10. $\sigma_\mathrm{RMS}$ is 1.5 \mjybeam, 35 \mujybeam, 60 \mujybeam and 30 \mujybeam for the LOFAR 54\,MHz (a), Apertif 1360\,MHz (b), VLA 1455\,MHz (c) and VLA 4700\,MHz (d) image, respectively. For the high resolution images, $\sigma_\mathrm{RMS}$ is 78 \mujybeam,  30 \mujybeam and 28 \mujybeam for the (e) LOFAR 144\,MHz, (f) VLA 4700\,MHz and (g) VLA 8415\,MHz image, respectively. The LOFAR 144\,MHz sub-arcsecond image has a lower resolution of 0.4\arcsec$\times$0.5\arcsec, than the one in Fig.~\ref{lofar_hba}. The negative contours are marked by dashed lines and are at $-3\sigma_\mathrm{RMS}$ level. Image statistics are summarised in Table~\ref{image_data}.}
        \label{continuum_lowres}
\end{figure*}
\end{appendix}
\end{document}